# Combined Thermal Expansion and Hydrolytic Stability Study of Lanthanide Vanadates $LnVO_4$ and $CaLnZr(VO_4)_3$ (Ln = La, Nd, Sm, Eu, Gd, Dy, Yb) with Zircon and Monazite Structures

A.K. Koryttseva[a][*], A.I. Orlova[a], N.S. Litonova[a], A.V. Nokhrin[a], M.S. Boldin[a], A.A. Murashov[a], D.G. Fukina[a], A.A. Atopshev[a], K.E. Smetanina[a], A.I. Beskrovnyi[b], V.A. Turchenko[b], N.Yu. Tabachkova[c, d]

[a] Lobachevsky State University of Nizhny Novgorod, 603022, Nizhny Novgorod, Russia

[b] Joint Institute for Nuclear Research, 141980, Dubna, Russia

[c] National University of Science and Technology "MISIS", 119049, Moscow, Russia

[d] A.M. Prokhorov General Physics Institute, Russian Academy of Science, 119049, Moscow, Russia

e-mail: koak@chem.unn.ru

**Abstract.** The paper presents the investigation of ordinary and ternary vanadates with zircon and monazite structures. Vanadates $LnVO_4$ and $CaLnZr(VO_4)_3$, where (Ln = La, Nd, Sm, Eu, Gd, Dy, Yb) and solid solution $La_{0.3}Nd_{0.5}Sm_{0.1}Eu_{0.1}VO_4$, were prepared by precipitation reaction. Bulk ceramic samples were obtained from powders by Spark Plasma Sintering (SPS). The powders and ceramics were examined with several physicochemical methods. The coefficients of thermal expansion at 900°C were determined. The hydrolytic stability of ceramic materials was studied. The peculiarities of high-rate SPS of powders of ordinary and ternary vanadates were analyzed.

**Abbreviations:** HLW – high-level radioactive waste, XRD (analysis) – X-ray diffraction (analysis), SEM – Scanning Electron Microscopy, TEM – Transmission Electron Microscopy, EDS (analysis) – Energy Dispersion (analysis), SPS – Spark Plasma Sintering, CXR – Characteristic X-ray radiation, CTE – Coefficient of thermal expansion.

---

[*] Corresponding Author (koak@chem.unn.ru)

# 1. Introduction

Synthetic compounds and natural minerals with monazite and zircon structures possess outstanding physical and chemical properties, such as high melting temperatures, high damage tolerance, high temperature resistance [1-6], optical properties [7-12], ionic and electronic conductivities [13-15], magnetic properties [16-19], chemical stability [20-22]. All these determines the interest in these structures compounds and its extensive research in different laboratories around the world. There are two most numerous fields of attention for the researchers: an applications as phosphors, lasers, light emitters and an application as matrix for radioactive waste management [23-30], in particular, *f*-elements of lanthanides, *etc*. The latter is due to the following features of their crystalline structures. Firstly, the centuries-long stability of the monazite and zircon structures under the conditions of different geological eras, what allows to predict similar stability of the synthesized analogues. Namely, the resistance to high temperatures, pressures, extreme long-term irradiation and other factors of both natural and synthetic origin should be taken into account [30-33].

Secondly, this structure has wide possibilities for iso- and heterovalent isomorphic substitutions, both in cationic and anionic sites of the structure. The cation sites could be inhabited by lanthanides, actinides, alkaline earth elements, tetravalent elements, *etc.* in various combinations and ratios. Thus, a wide variety of compositions could crystallizes in the monazite and zircon structures yielding a huge family of phosphates, silicates, arsenates, chromates, sulfates, vanadates. The stability limits of these crystalline forms regarding compositions, temperatures, and pressures correlate well with crystallographic criteria and described in details in reviews [32, 33] and other key articles.

The most of the publications relate to the synthesis and study of ordinary Ln phosphates with the monazite and zircon structures [32]. The cooperative incorporation of elements in +2, +2, +4 oxidation states in these structures with isomorphic substitution of anionic groups, for example, phosphate ions with arsenate, vanadate, chromate ions, etc., is much less studied [32, 34-35].

In this work, two series of Ln-containing vanadates were prepared and investigated. The first contains Ln and vanadate ion in an equimolar ratio and have a formula LnVO$_4$, we assigned them as "ordinary" vanadates. In the second group, cations in +2 and +4 oxidation states are incorporated together with Ln$^{3+}$ into the structure. The following cationic substitution takes place 3Ln$^{3+}$ → Ln$^{3+}$ + Ca$^{2+}$ + Zr$^{4+}$, leading to the formula CaLnZr(VO$_4$)$_3$. We assigned them as "ternary" vanadates. Two groups was selected to compare crystallographic and other characteristics of compounds belonging to the same structural type, but differing in mole contents of lanthanides in their composition. Ordinary Ln vanadates contain 20 mol.% Ln, whereas ternary ones contain 5.6 mol.%. A solid solution of complex cationic composition La$_{0.3}$Nd$_{0.5}$Sm$_{0.1}$Eu$_{0.1}$VO$_4$ was also synthesized.

The compounds were prepared by different methods to find out a connection between the synthesis method, the microstructure and the sintering behavior of vanadate powders. The possibilities of powders consolidation into ceramic materials using the Spark Plasma Sintering (SPS) [36-40] method have been studied. The SPS is a promising technology process that produced of highly-stable mineral-like ceramics for immobilization of high-level components in radioactive waste (HLW), minor actinides transmutation and Nuclear Fuel Cycle [37, 41-47]. Thermal expansion and hydrolytic stability of the two above-mentioned series of vanadates was studied. All obtained characteristics obtained were discussed from the point of possible proposal of these vanadates as matrices for the immobilization of radioactive waste.

## 2. Materials and Methods

### 2.1 Synthesis and sintering

**Starting Materials** (of chemically pure grade) as lanthanum oxide La$_2$O$_3$, neodymium oxide Nd$_2$O$_3$, samarium oxide Sm$_2$O$_3$, europium oxide Eu$_2$O$_3$, gadolinium oxide Gd$_2$O$_3$, dysprosium oxide Dy$_2$O$_3$, ytterbium oxide Yb$_2$O$_3$, ammonium metavanadate NH$_4$VO$_3$, zirconyle oxy-chloride ZrOCl$_2$×8H$_2$O, calcium carbonate CaCO$_3$, urea (NH$_2$)$_2$CO, citric acid (C$_6$H$_8$O$_7$×H$_2$O) were used with no additional refinement.

All compounds were prepared by precipitation reaction with ammonium metavanadate. Vanadates LnVO$_4$, where Ln = La, Nd, Sm, Eu, Gd, Dy, Yb and La$_{0.3}$Nd$_{0.5}$Sm$_{0.1}$Eu$_{0.1}$VO$_4$, were prepared according to the procedure, reported in [48, 49]. Stoichiometric amounts of Ln$_2$O$_3$ and NH$_4$VO$_3$ were dissolved in nitric acid (6N solution), then the excess amount of urea was added, and mixture was gradually heated to 400 °C and the melt was evaporated.

The synthesis of ternary vanadates CaLnZr(VO$_4$)$_3$, where Ln = Nd, Sm, Eu, Gd, Dy, Yb, was carried out by two routes: synthesis in molten urea, described above (synthesis #1) and synthesis in a solution with a complexing agent (synthesis #2).

In synthesis #2, stoichiometric quantities of lanthanide oxides, calcium carbonate, zirconium oxychloride hydrate were dissolved in nitric acid (1N solution), then the excess amount of citric acid and a stoichiometric amount of precipitant, ammonium metavanadate NH$_4$VO$_3$, were added. Powder precursors from both methods were heat treated: at 600, 700, 800, 900 °C for 4-8 h, alternating with dispersion in an agate mortar after each stage.

The mass of the prepared powders was varied from 0.3 to 1g. The repeatability of the results was studied by repeating synthesis no less than 5 times.

**Ceramics preparation**

Dr. Sinter model SPS-625 (SPS SYNTEX, Japan) setup was used for ceramic preparation. Temperature was measured using a Chino IR-AHS2 optical pyrometer focused on the outer surface of the graphite mold. The accuracy of temperature determination was ± 20 °C. Heating of the powder material was carried out by passing sequences of high-power direct current pulses (up to 5 kA). Sintering was carried out in graphite molds with an internal diameter of 10.8 mm. To increase the density of adhesion of the powder to the inner surface of the mold, graphite foil was used. Sintering was carried out in vacuum, with the application of uniaxial pressure (45-50 MPa). Two-stage heating was used: Stage #1 – heating at a rate of 100 °C/min to a temperature of 600 °C; Stage #2 – heating at a rate of 50 °C/min to a temperature of 800 °C. There was no exposure at a temperature of 800 °C.

The samples were cooled together with the installation. At the cooling stage, no pressure was applied to avoid sample destruction.

The effective powders shrinkage ($L_{eff}$) was measured using a Futaba Pulscale SMM151A dilatometer attached to the Dr. Sinter model SPS-625. Empty mode heating were performed to take into account the contribution of thermal expansion of the mold ($L_0$), an experiment was carried out on heating empty molds. True shrinkage (L) was determined by the formula: $L(T) = L_{eff}(T) - L_0(T)$. The accuracy of shrinkage determination was ± 0.01 mm. The contribution of thermal expansion is important when sintering samples with low relative density (low powder shrinkage on heating) at low temperatures.

Annealing of ceramic samples to eliminate graphite contamination from ceramic samples [50] was carried out in an air atmosphere in an EKPS-10 furnace at a temperature of 700 °C for 2 hours. Heating and cooling of the samples was done together with the furnace.

2.2 Samples characterization

**X-ray diffraction study.** Powder X-ray diffraction (XRD) was performed on an XRD-6000 diffractometer (Shimadzu LabX) in the 2θ angle range from 10 to 50° at a rate of 2 deg/min at an X-ray wavelength $\lambda_{Cu}$ = 1.54056 Å. The presence of impurity phases in powders and ceramics was monitored using an XRD-7000 diffractometer (Shimadzu, Japan). Scanning was carried out in the angle range 17-67°, scanning step 0.04°, exposure time 2 s. Bragg-Brentano focusing was used. Qualitative phase analysis was carried out in the DIFFRAC.EVA program (Bruker) using data from the PDF-2 bank [51]. Quantitative phase analysis and calculation of unit cell parameters were carried out by the Rietveld method in the Topas (Bruker) software package. The average error in determining the mass content of phases was 0.2%.

**Electron microscopy.** Transmission electron microscopy (TEM) were used together with scanning electron microscopy (SEM). The combined use of two methods made it possible to study powders that were opaque to an electron beam. For example, Gd-containing samples were studied by SEM.

The microstructure and morphology of powders were studied using a Tescan Vega 2 SEM and a Jeol JEM-2100F TEM. The element composition of the sample was studied by Energy Dispersive X-ray (EDS) microanalysis with an X-MaxN 20 detector (Oxford Instruments): Kα(O, Ca, V) and Lα(Zr, Sm, Nd, Yb). The X-Man20 detector is attached with JSM-IT300LV SEM. The error of heavy elements determination is 0.2-0.5 at.%, and error for light elements - 1-2 at.%.

**High temperature XRD.** High-temperature (25-900°C) XRD studies were carried out with an Anton Paar HTK 1200N X-ray camera installed on an Empyrean PANalytical diffractometer using CuKα radiation. Scanning was performed using Bragg–Brentano focusing, with a scanning step of $\Delta(2\theta) = 0.023°$. The exposure time at each point was 240 s. The powders were poured into a 3 mm deep cuvette and manually compacted, poured again and compacted until the surface of the powder reached the cuvette edge. This procedure was necessary to maintain the maximum diffractometer resolution. The accuracy of temperature maintaining in the chamber is ±0.1 degrees. After reaching the specified temperature, the sample was kept for 10-15 min.

**Density** ($\rho$) was measured by hydrostatic weighing using a Sartorius CPA 225D balance. The error of the relative density $\rho/\rho_{th}$ determination was ± 0.2%, the absolute density – 0.05 g/cm$^3$.

**The study of hydrolytic stability** was carried out according to the Russian National Standard GOST R 52126-2003 (analogous to the ASTM C1220-21 standard), with the MCC-1 method [52]. Ceramic sample was thoroughly washed and placed into a container (flat-bottomed flask) for leaching test, then filled with a contact solution of known volume. Contacting water was replaced after 1, 2, 4, 7, 9, 11, 14, 21, 28 days from the start of the experiment. At the specified time, the samples were removed from the container and washed out with a fresh portion of contacting solution of a volume equal to the volume of the contacting solution. The washing solution was added to the spent contact solution. The sample, without allowing it to dry, was placed in the same container and filled with a new portion of the contact solution. The tests were carried out at room temperature; double distilled water was used as a contact solution. To calculate the rate of leaching of cations from ceramic samples, initially we calculated the mass loss of *i* component using the formula:

$$NL = \frac{a_{ki} \cdot m_{обр.}}{a_{0i} \cdot S}, \qquad (1)$$

where $a_{0i}$ – mass of *i* component in the sample of $m_{обр}$ mass, $a_{ki}$ – mass of *i* component, passed into solution during the leaching process, g; $S$ – sample surface area, sm$^2$.

Leaching rate of *i*-component ($R_i$, g·sm$^{-2}$·days$^{-1}$) was determined by the formula:

$$R_i = \frac{dNL}{dt},$$

where t – duration of leaching period, days. The ion concentration was determined by inductively coupled plasma atomic emission spectrometry.

### 3. Results

3.1 XRD phase analysis and electron microscopic studies

Figures 1-2 show XRD patterns of the synthesized vanadates. XRD results indicate the formation at 800 and 900 °C of monophase products LnVO$_4$, where Ln = La, Nd, Sm, Eu, Gd, Dy, Yb; monophase solid solution La$_{0.3}$Nd$_{0.5}$Sm$_{0.1}$Eu$_{0.1}$VO$_4$ and monophase vanadates CaLnZr(VO$_4$)$_3$, where Ln = Nd, Sm, Eu, Gd, Dy, Yb. All compounds (except for LaVO$_4$) are characterized by tetragonal syngony and belong to the zircon type structure (Ca$_{0.5}$Zr$_{0.5}$VO$_4$ analogue (PDF #01-070-2000, ICSD #6111), space group I4$_1$/amd). The LaVO$_4$ compound crystallizes in the monazite type structure (LaPO$_4$ analogue, space group P2$_1$/n). Unit cell parameters of all vanadates increase with increasing ionic radius of the lanthanide [53] (Fig. 3-4). The results obtained agree well with data on other vanadates and phosphates with zircon type structure [32]. Ternary vanadates show less pronounced dependence of the lattice parameters on the ionic radius than for simple LnVO$_4$ vanadates, what is caused by less mole % of Ln per formula unit.

CaEuZr(VO$_4$)$_3$, CaSmZr(VO$_4$)$_3$ and CaDyZr(VO$_4$)$_3$ powders prepared from urea melt, have a small amount of the impurity phase of zirconium oxide m-ZrO$_2$ (PDF #00-036-0420, ICSD #57157). Whereas the amount of the m-ZrO$_2$ impurity reaches 1.6-1.7 wt.% for CaEuZr(VO$_4$)$_3$ and CaSmZr(VO$_4$)$_3$ samples prepared with citric acid. The reflexes corresponding to the m-ZrO$_2$ only slightly exceed the background level on the XRD patterns of CaDyZr(VO$_4$)$_3$ powders, and it is impossible to reliably determine the impurity content in these powders using the XRD method. We conclude, that changing the preparation method does not have a noticeable effect on the m-ZrO$_2$ impurity amount in CaDyZr(VO$_4$)$_3$ powders.

Fig. 5 presents SEM results for ternary vanadate powders CaLnZr(VO$_4$)$_3$. As an example Fig. 5a, b show images of vanadate powders CaYbZr(VO$_4$)$_3$, and in Fig. 5c, d – vanadate powders CaNdZr(VO$_4$)$_3$. There are single agglomerates up to 20-40 µm in size, consisting of individual particles of various sizes and shapes in all CaLnZr(VO$_4$)$_3$ powders. CaYbZr(VO$_4$)$_3$ powders have a faceted shape, the particle size varies over a wide range - from nano- and submicron particles to particles whose size reaches 4-5 µm. Powders contain elongated particles with the length of several times greater than width. The shape of CaNdZr(VO$_4$)$_3$ particles (Fig. 5c, d) is close to that for CaYbZr(VO$_4$)$_3$ particles, but CaNdZr(VO$_4$)$_3$ powders practically have no small submicron particles, which are characteristic ones for CaYbZr(VO$_4$)$_3$ powders .

LnVO$_4$ particles have a micron size and shape close to spherical. There are practically no small nano- and submicron particles in powder samples of simple vanadates LnVO$_4$. Fig. 6 shows electron microscopic images of SmVO$_4$ (Fig. 6a, b) and GdVO$_4$ (Fig. 6c, d) powders.

Attention should be paid to the difference in the morphology of CaYbZr(VO$_4$)$_3$ powders prepares by two different methods (Fig. 7). Regardless of the preparation route, CaYbZr(VO$_4$)$_3$ powders are highly agglomerated, but the use of citric acid leads to a decrease in the agglomerates size (less than 5 µm) and to a decrease in particles size inside the agglomerates (less than 1 µm). Denser agglomerates with sizes up to 5 µm or more are formed in the case of synthesis in molten urea, and here polycrystals have sizes of about 1 µm. However, the agglomerates size is small and does not significantly affect the parameters of the sintered ceramic. Particles of CaYbZr(VO$_4$)$_3$ powders obtained from a solution using citric acid contain a large number of dislocations (Fig. 7c, d), which form a significant elastic stress fields around themselves. Whereas, particles with an increased dislocation density were not practically observed (Fig. 7e, f) in powders prepared from molten urea.

The TEM results indicate a thin amorphous layer of several nanometers thick (Fig. 8a) on the surface of the powders. We suppose, that such a layer might be one of the reasons of the submicron agglomerates formation from several nanoparticles (Fig. 8b) and also formation of larger agglomerates of several microns in size (Fig. 8c).

The EDS study showed that the elements ratio corresponds to the expected ratio within the sensitivity limits of the method (1-3 wt.% depending on the chemical element) for most powder compounds ($SmVO_4$, $NdVO_4$, $CaNdZr(VO_4)_3$, $CaYbZr(VO_4)_3$, etc.). An additional error increase in the determination of vanadium and oxygen is due to the superposition of their lines $K\alpha(O) = 0.525$ eV and $K\alpha(V) = 0.511$ eV by more than 40%. An overestimation of the Zr content is observed (from 0.5 to ~ 2 at.% for different samples) for $CaNdZr(VO_4)_3$ and $CaYbZr(VO_4)_3$ powders, which is probably due to the presence of an zirconium oxide impurity, and is confirmed by the uneven distribution of Zr on element distribution maps (Table 1).

**Table 1**. Results of EDS analysis of the elemental composition of samples

| Title formula | Sample type | Experimental composition (at. %) | | | | | | |
|---|---|---|---|---|---|---|---|---|
| | | Ca (Kα) | Sm (Lα) | Yb (Lα) | Nd (Lα) | Zr (Lα) | V (Kα) | O (Kα) |
| $SmVO_4$ | ceramic | - | 18.90 | - | - | - | 19.1 | 62.0 |
| $NdVO_4$ | ceramic | - | - | - | 21.2 | - | 24.0 | 54.8 |
| $CaNdZr(VO_4)_3$ | ceramic | 6.2 | - | - | 5.1 | 8.8 | 17.0 | 62.9 |
| $CaNdZr(VO_4)_3$ | powder | 6.3 | - | - | 5.8 | 6.8 | 19.4 | 61.7 |
| $CaYbZr(VO_4)_3$ | ceramic | 6.8 | - | 6.6 | - | 7.7 | 19.6 | 59.3 |
| $CaYbZr(VO_4)_3$ | powder | 6.2 | - | 5.1 | - | 6.5 | 17.4 | 64.8 |

*In situ* XRD in the temperature range (25-900°C) showed that, LnVO$_4$ compounds maintain their phase composition, exhibiting high resistance to heating. Fig. 9 shows XRD patterns for SmVO$_4$ powder samples. Analysis of the presented high-temperature XRD patterns confirms that heating does not lead to the appearance of any new phases. It can also be seen that as the heating temperature increases, the width of the maxima and their XRD peak intensity do not change. This also indicates the high stability of the zircon vanadate structure on heating.

CaLnZr(VO$_4$)$_3$ compounds have less thermal stability of the structure and when they are heated, a small amount of impurity phases appears. Figs. 10 and 11 show high-temperature XRD patterns obtained during *in situ* studies of CaYbZr(VO$_4$)$_3$ (Fig. 10) and CaNdZr(VO$_4$)$_3$ (Fig. 11). Areas of the XRD patterns which contain changes that occur on heating of the powder are highlighted on Figs. 10-11. Green rectangle (Fig.10) highlights the areas (25-530 °C) of the XRD patterns of the CaYbZr(VO$_4$)$_3$, where the maxima of the YbVO$_4$ phase are present. The red rectangle in Fig. 10 represent a (480-530 °C) section, which contains maximum related to the ZrO$_2$ tetragonal modification. The area (725-900 °C) containing, presumably, the maximum of the monoclinic modification of ZrO$_2$ (Fig. 10) is highlighted with a black rectangle. Despite the low intensity of the above discussed impurities maxima they are visible and clearly distinguishable on the XRD patterns. We also note, that, the width of the maxima decreases and their intensity increases with increasing heating temperature of the CaYbZr(VO$_4$)$_3$ powder, which indicates the particle growth or the defects annealing. Fig. 11 shows high-temperature XRD pattern of CaNdZr(VO$_4$)$_3$ powder. The main phase belongs to zircon structure type (Ca$_{0.5}$Zr$_{0.5}$VO$_4$ analogue). A black rectangle on Fig. 11 highlights the area corresponding to the V$_4$O$_7$ phase, which is observed on all diffraction patterns. Similar results were obtained when analyzing the *in situ* results of the remaining compounds. Thus, the XRD results indicate the possibility of the impurity formation in the vanadates under study. Also we emphasize that in most cases the impurity peaks have very low intensity (Fig. 10, 11), indicating the content of impurity phases to be quite small and they do not have a significant effect on the temperature behavior of the target CaLnZr(VO$_4$)$_3$ phase.

The axial ($α_a$, $α_b$, $α_c$) and average coefficients ($α_{av}$) of thermal expansion (CTE) were calculated (Table 2), based on the analysis of the temperature dependence of unit cells parameters of the target phase (Fig. 12, 13). The results obtained indicate that these vanadates belong to the medium-expanding substances. The vanadate solid solution $La_{0.3}Nd_{0.5}Sm_{0.1}Eu_{0.1}VO_4$ obtained in this work with a zircon structure has a lower CTE ($α_{av} = 6·10^{-6}$ K$^{-1}$) compared to the solid solution of phosphate $La_{0.3}Nd_{0.5}Sm_{0.1}Eu_{0.1}PO_4$ with the structure monazite ($α_{av} = 9.64·10^{-6}$ K$^{-1}$), which was obtained earlier [54].

**Table 2.** Axial and average coefficients of thermal expansion of crystal unit cell of vanadates $LnVO_4$ [5] and $CaLnZr(VO_4)_3$ (30-900°C)

| Compound | $α_a·10^6$, K$^{-1}$ | $α_b·10^6$, K$^{-1}$ | $α_c·10^6$, K$^{-1}$ | $α_{av}·10^6$, K$^{-1}$ |
|---|---|---|---|---|
| Monazite structure | | | | |
| $LaVO_4$ [5] | 9.16 | 6.08 | 13.9 | 9.74 |
| Zircon structure | | | | |
| $NdVO_4$ [5] | 3.44 | 3.44 | 10.3 | 5.73 |
| $SmVO_4$ [5] | 3.52 | 3.52 | 10.29 | 5.77 |
| $EuVO_4$ [5] | 4.15 | 4.15 | 10.3 | 6.23 |
| $GdVO_4$ [5] | 4.05 | 4.05 | 10.1 | 6.06 |
| $DyVO_4$ [5] | 3.58 | 3.58 | 10.05 | 5.89 |
| $YbVO_4$ [5] | 4.35 | 4.35 | 10.1 | 6.29 |
| $CaNdZr(VO_4)_3$ | 4.07 | 4.07 | 13.04 | 7.06 |
| $CaSmZr(VO_4)_3$ | 4.21 | 4.21 | 13.06 | 7.16 |
| $CaEuZr(VO_4)_3$ | 3.22 | 3.22 | 12.4 | 6.29 |
| $CaGdZr(VO_4)_3$ | 4.28 | 4.28 | 13.1 | 7.24 |
| $CaDyZr(VO_4)_3$ | 3.45 | 3.45 | 11.3 | 6.102 |
| $CaYbZr(VO_4)_3$ | 4.15 | 4.15 | 12.5 | 6.96 |

The decomposition of the compounds under study begins at higher heating temperatures (more than 900 °C). According to XRD data [49], after annealing at 1000 °C, the compound CaZrYb(VO$_4$)$_3$ decomposes into simple vanadates and oxides:

$$3CaYbZr(VO_4)_3 = 3YbVO_4 + Ca_3(VO_4)_2 + 3ZrO_2 + 2V_2O_5.$$

Similar results were obtained for other compounds. This became the reason why the maximum sintering temperature of ceramics was limited.

3.2 Ceramics preparation and characterization

Ceramic samples were prepared from all LnVO$_4$ and CaLnZr(VO$_4$)$_3$ powders using the SPS method. A typical diagram of ceramic sintering "Temperature (T) – Applied Pressure (P) – Vacuum Pressure (V$_{ac}$) – Time (t)" is shown in Fig. 14. There is a slight decrease in vacuum pressure during ceramic heating (Fig. 14), which is not observed when heating an empty graphite mold in a vacuum. The time curve of vacuum pressure for heating an empty graphite mold is indicated as V$_{ac}$(2) on Fig. 14. Note also that the vacuum pressure in the case of heating ceramics is less than the vacuum pressure in the case of heating an empty graphite mold: V$_{ac}$ < V$_{ac(2)}$. We assume, that a vacuum pressure decrease indirectly indicates a slight dissociation of elements from the surface of ceramic samples. The maximum heating temperature determined by the Chino IR-AHS2 pyrometer was limited to 800 °C. The 800 °C limitation was due to the fact that the actual surface temperature of the sample during SPS turns out to be higher than the temperature, determined by a pyrometer, focused on the surface of the graphite mold [55]. The ceramics sintering modes together with the achieved densities are collected in Table 3.

**Table 3.** SPS parameters of vanadate powders

| № | Compound | $T_s$, °C | P, MPa | $V_h$, °C/min | $\tau_s$, min | $\rho_{th.}$, g/sm$^3$ | $\rho/\rho_{th}$, % | $mQ_s$, $kT_m$ |
|---|---|---|---|---|---|---|---|---|
| 1 | LaVO$_4$ | | | | | 3.2532 | 64.2 | 2.6 |
| 2 | NdVO$_4$ | | | | | 3.6412 | 73.2 | 2.6 |
| 3 | SmVO$_4$ | | | | | 4.0672 | 77.5 | 3.0 |
| 4 | EuVO$_4$ | | | | | 4.1340 | 77.9 | 7.3 |
| 5 | GdVO$_4$ | | | | | 3.1219 | 56.9 | 3.2 |
| 6 | CaNdZr(VO$_4$)$_3$ | 800 | 50 | 50 | 0 | 3.3554 | 58.0 | 4.8 |
| 7 | CaSmZr(VO$_4$)$_3$ | | | | | 3.1515 | 53.9 | 4.4 |
| 8 | CaEuZr(VO$_4$)$_3$ | | | | | 3.3768 | 57.6 | 4.6 |
| 9 | CaGdZr(VO$_4$)$_3$ | | | | | 3.3955 | 57.3 | 4.9 |
| 10 | CaDyZr(VO$_4$)$_3$ | | | | | 3.7828 | 61.4 | 7.8 |
| 11 | CaYbZr(VO$_4$)$_3$ | | | | | 4.0132 | 64.2 | 8.0 |
| 12 | La$_{0.3}$Nd$_{0.5}$Sm$_{0.1}$Eu$_{0.1}$VO$_4$ | | | | | 1.9788 | 40.0 | -$^{(*)}$ |

$^{(*)}$ the reliability of the linear approximation of the L(T) curve in the coordinates $\ln(T\partial\varepsilon/\partial T) - T_m/T$ is very low $R^2 < 0.5$ (see section Discussion); it is impossible to reliably determine the value of the SPS effective activation energy $mQ_s$

Most of the samples looked like loose compacts, due to their low relative density (Table 3). The relative density of LnVO$_4$ ceramics did not exceed 78%, and relative density of CaLnZr(VO$_4$)$_3$ ceramics did not exceed 65%. The relative density of La$_{0.3}$Nd$_{0.5}$Sm$_{0.1}$Eu$_{0.1}$VO$_4$ ceramics was 40%; the sample was easily fractured under mechanical stress. Repeated heating up to a higher temperature increases the density of ceramic samples, but leads to the fracture of some compact samples and to their chemical decomposition (see Appendix A).

According to XRD data, the structure of ceramic samples maintains the structure of the initial powders. The low heating temperature made it possible to avoid the process of decomposition of the compound and minimize the appearance of impurities in most of the compounds under study. A slight increase of the $V_4O_7$ phase content is observed (card #04-005-4524 in the PDF-4 database, card #01-071-0423 in the PDF-2 database) for $SmVO_4$ ceramics, as an exception. However, the intensity of the $V_4O_7$ XRD peak does not exceed three times background level in the diffraction pattern. The appearance of a monoclinic mon-$ZrO_2$ (PDF 00-036-0420), which was initially absent in the starting powder, was detected in $CaYbZr(VO_4)_3$ ceramics. All other $CaLnZr(VO_4)_3$ powders show no increase in the mon-$ZrO_2$ impurity amount.

Fig. 15-16 presents the results of SEM studies of the microstructure of sintered ceramics. Ceramics have a rather loose microstructure; the particle size in ceramics is close to the particle size in the original powders. The volume fraction of small submicron particles, which were initially present in large quantities in $CaLnZr(VO_4)_3$ powders, decreases significantly. The presence of particles with significantly different sizes in the initial powders simultaneously leads to the appearance of a ununiform (heterogeneous) microstructure in the ceramics (Fig. 16). Areas with large grains are highlighted with a dotted line (Fig.16).

The results of EDS analysis of the ceramics composition are presented in Table 1 and in Fig. 15. Fig. 15 shows that most ceramics have a uniform distribution of chemical elements over the sample surface. Local changes in the composition of ceramics are associated with the presence of pores on the sample surface and the appearance of corresponding artifacts on SEM images (f. e., for $NdVO_4$ ceramics on Fig. 15b). Quite large $ZrO_2$ particles are clearly visible on the surface of some $CaLnZr(VO_4)_3$ ceramics. Appearance or increase of the $ZrO_2$ amount during $CaYbZr(VO_4)_3$ sintering is observed on the ceramics surface (Fig. 15d) (see text above). These impurity particles became clearly visible on the map of zirconium distribution over the sample surface. Table 1 shows the composition of the ceramics to be close to the composition of the starting powders within the sensitivity of the EDS method.

Table 4 and Fig. 17 presents the results of hydrolytic tests of ceramics. The minimum achieved leaching rates of lanthanide and vanadium ions are given in Table 4. Typical graphs of normalized mass loss $NL_i$ and leaching rate $R_i$ versus time $\tau$ of some ceramic samples are plotted in Fig. 17 for some selected ceramic samples. Fig. 17 shows that $NL_i(t)$ and $R_i(t)$ dependences have a classical monotonic character, while the $NL_i$ and $R_i$ values quickly reach their stationary values for $LnVO_4$ ceramic samples. Whereas, $NL_i$ and $R_i$ values do not reach their stationary values even on the $28^{th}$ day of testing for ceramic samples $CaLnZr(VO_4)_3$ and $Ln_{0.3}Nd_{0.5}Sm_{0.1}Eu_{0.1}VO_4$.

**Table 4.** Lanthanum and Vanadium leaching rates from ceramic vanadate samples. 28 days static test

| Compound | Leaching rate (Ln) $R_{min} \cdot 10^8$, g·sm$^{-2}$·d$^{-1}$ | Leaching rate (V) $R_{min} \cdot 10^5$, g·sm$^{-2}$·d$^{-1}$ |
|---|---|---|
| $LaVO_4$ | 0.04 | 0.03 |
| $NdVO_4$ | 0.08 | 0.64 |
| $SmVO_4$ | 0.15 | 0.32 |
| $EuVO_4$ | 0.13 | 0.27 |
| $GdVO_4$ | 0.04 | 0.02 |
| $DyVO_4$ | 0.04 | 0.04 |
| $YbVO_4$ | 0.06 | 0.03 |
| $CaNdZr(VO_4)_3$ | 0.19 | 1.56 |
| $CaSmZr(VO_4)_3$ | 0.31 | 0.54 |
| $CaEuZr(VO_4)_3$ | 0.11 | 0.51 |
| $CaDyZr(VO_4)_3$ | 0.12 | 0.70 |
| $CaYbZr(VO_4)_3$ | 0.24 | 0.70 |
| $Ln_{0.3}Nd_{0.5}Sm_{0.1}Eu_{0.1}VO_4$ | 0.42; 0.26; 0.22; 0.95 | 0.35 |

Ceramic vanadates samples maintained their integrity (with the exception of vanadate $Ln_{0.3}Nd_{0.5}Sm_{0.1}Eu_{0.1}VO_4$, which initially had a very low relative density) after 28 days of hydrolytic test duration. The formation of large secondary phases or hardness salts on the surface of ceramic samples after hydrolytic tests is not observed. Table 4 shows that the leaching rate of the most representative components in water, which is no more than ~$10^{-5}$–$10^{-6}$ g·cm$^{-2}$·d$^{-1}$· day, meets the requirements for materials for radioactive waste disposal and is comparable with other potential matrices [56].

According to XRD data, it is clear that the phase composition of the ceramics did not change after hydrolytic tests (Fig. 18). Fig. 18 shows that the diffraction patterns of the samples after hydrolytic tests contain all the same XRD peaks, which were present in the diffraction patterns of the samples before hydrolytic tests. The results obtained allow to assume that during the hydrolytic tests there is no significant destruction of the zircon vanadate structure.

## 4. Discussion

First, let us compare the data obtained with known literature data on other ordinary and ternary lanthanide phosphates and vanadates. Ordinary lanthanide phosphates crystallize in the zircon structure for lanthanides (Gd-Lu) and in the monazite structure for larger in size (lighter) lanthanides (Ce-Dy). Gd, Tb, Dy phosphates exist in 2 polymorphs: monazite and zircon. Ordinary Ln vanadates have a zircon structure for all lanthanides, with LaVO$_4$ being the only orthovanadate that exists in both structures (monazite and zircon). This analysis shows that the type of crystal structure is determined primarily by the sizes of the structure forming anions. As the anion size increases, the compounds preferably crystallize in the zircon structure. In our case, all obtained vanadates crystallize in the zircon structure, regardless of the lanthanide radius. This differs them from similar orthophosphates CaGdZr(PO$_4$)$_3$ and CaEuZr(PO$_4$)$_3$, which crystallize in the monazite structure [32].

Thus, the experimental data obtained expand our knowledge on the isomorphism in the zircon structure. The isomorphic replacement of the phosphate ion PO$_4^{3-}$ by the vanadate ion PO$_4^{3-}$ leads to the crystallization of larger lanthanides in the zircon structure. As the anion size increases, zircon unit cell volume increases, allowing larger lanthanides to adopt this structure in contrast with phosphates, crystallizing in monazite structure with small 2-, 3-, and 4-valent elements.

Let us now analyze the features of high-speed sintering of ceramic vanadates $LnVO_4$ and $CaLnZr(VO_4)_3$, where (Ln = La, Nd, Sm, Eu, Gd, Dy, Yb). Fig. 19 shows the effective temperature shrinkage curves $L_{eff}(T)$ for vanadates $LnVO_4$ (Fig. 20a) and $CaLnZr(VO_4)_3$ (Fig. 19b). Further, we will not analyze the $L_{eff}(T)$ curves for the vanadates $GdVO_4$ and $LaVO_4$, since their shrinkage is small and it is not possible to reliably analyze the dependences obtained. The contribution of thermal expansion $L_0(T)$ during SPS in this temperature range is quite large and, therefore, the main contribution to the nature of the true shrinkage curve L(T) in the case of $GdVO_4$ and $LaVO_4$ vanadates is made by the thermal expansion curve of the graphite mold $L_0(T)$.

Fig. 19a, b clearly show that the $L_{eff}(T)$ curves have a three-stage character. The first stage of intense shrinkage is observed in the region of low heating temperatures (Stage I). For ternary vanadates, the first stage of intense shrinkage is observed in the temperature range of 400-500 °C; for ordinary vanadates, the temperature of the Stage I beginning is ~220-250 °C. In the region of medium heating temperatures (~500-600 °C), a stationary stage is observed on the L(T) curves, where the shrinkage of vanadate powders practically does not change. This stage is most clearly seen in the L(T) curves for ordinary vanadates $EuVO_4$ and $SmVO_4$ (Fig. 19c), and also for almost all vanadates $CaLnZr(VO_4)_3$ (except for $CaEuZr(VO_4)_3$) (Fig. 19d). At heating temperatures above 600 °C, a second stage of intense shrinkage is observed for ordinary and ternary vanadate powders. We suppose the change in the shrinkage curves nature to be related to, first of all, the heating rate change at a temperature of 600 °C: heating from room temperature to 600 °C was carried out at a rate of 100 °C/min, heating from 600 to 800 °C was carried out at a rate of 50 °C/min. A change in the heating rate leads to a change in the slope of the $L_0(T)$ curve and also in the shrinkage rate of the powders.

Analysis of the powders compaction kinetics at the "high temperature" stage of intensive compaction (Stage III) can be carried out using the Young-Cutler model [57]. This model describes the initial stage of non-isothermal sintering of spherical particles under conditions of simultaneous processes of volumetric and grain-boundary diffusion, as well as plastic deformation. According to [57], the dependence of shrinkage on heating temperature can be described using the equation:

$$\varepsilon^2(\partial\varepsilon/\partial t) = (2{,}63\gamma\Omega D_v\varepsilon/kTd^3) + (0{,}7\gamma\Omega bD_b/kTd^4) + (Ap\varepsilon^2 D/kT), \quad (1)$$

где $\varepsilon$ – shrinkage, t – time, $\gamma$ - free energy, $D_v$ - volume diffusion coefficient, $D_b$ - grain boundary diffusion coefficient, d – grain size, p – applied pressure, D – diffusion coefficient during plastic deformation. The slope angle of the $\ln(T\partial\varepsilon/\partial T)$ – $T_m/T$ dependence corresponds to the effective activation energy of the sintering process $mQ_{s2}$, where $m$ is a coefficient depending on the diffusion mechanism, $T_m$ is the melting temperature of the material (for the $NdVO_4$ compound – $T_m \sim 1700$ °C). According to [57], the $m$ coefficient could be of different values: $m = 1/3$ for the grain boundary diffusion, $m = 1/2$ - for diffusion in a crystal lattice, $m = 1$ - for viscous flow (creep) of the material. The effectiveness of the Young-Cutler model application for analysis the temperature shrinkage curves, obtained during SPS is shown in [55, 58-62].

Fig. 20 shows a typical $\ln(T\partial\varepsilon/\partial T)$ – $T_m/T$ curve for vanadates as an example. These dependencies can be described by two straight lines, with different slopes. According to [57], we can assume that the compaction initial stage with low activation energy is associated with the sliding of vanadate particles relative to each other under the applied uniaxial pressure. The intensity of the particles sliding relative to each other depends on their size, shape and initial compaction density. When particles reach their critical packing density, their sliding relative to each other becomes difficult, and the intensive powder shrinkage stops. With a further temperature increase, diffusion intensifies, leading to an exponential increase in the powder shrinkage rate.

The effective SPS activation energy for the stage of high-temperature intensive shrinkage are given in Table 3 for vanadate powders. The average error in the $mQ_s$ determination was about ± 0.5 $kT_m$. The effective SPS activation energy reaches 7.8-8.0 $kT_m$ for vanadates $CaDyZr(VO_4)_3$ and $CaYbZr(VO_4)_3$. For $LnVO_4$ vanadates, the effective SPS activation energy is close to 2.6-3.2 $kT_m$. Comparison of $mQ_s$ values with the data on Fig. 3, 4 shows that an increase in the unit cell volume of the vanadate crystal lattice leads to a decrease in $mQ_s$. For example, the effective SPS activation energy of vanadate $CaYbZr(VO_4)_3$ with a unit cell volume V ~ 306 Å$^3$ is ~8.0 $kT_m$, and that of vanadate $CaNdZr(VO_4)_3$ with V ~ 318 Å$^3$ is $mQ_s$ ~ 4.8 $kT_m$. We also note that ternary vanadates $CaLnZr(VO_4)_3$ have higher values of the effective SPS activation energy $mQ_s$ than ordinary vanadates $LnVO_4$. We suppose, this fact to be due to larger unit cell volume of $LnVO_4$ vanadates compared to unit cell volume of $CaLnZr(VO_4)_3$. The dependence of the effective SPS activation energy on the volume of the unit cell volume is presented in Fig. 21.

Fig. 21 and Table 3 shows that the exception is observed for the compound $EuVO_4$. This compound simultaneously has a fairly large unit cell volume (V ~ 334 Å$^3$) and a high sintering activation energy $mQ_s$ ~ 7.3 $kT_m$. At present, the reasons for the anomalous behavior of the $EuVO_4$ have not been determined.

At $m$ = 1/3, the obtained values of SPS activation energy of ternary vanadates $CaLnZr(VO_4)_3$ are lower than typical values of activation energy of volume diffusion [63] and are close to typical values of activation energy of grain boundary diffusion in ceramics [64]. This conclusion also corresponds well to the studies of fine-grained materials sintering, which have a large volume fraction of grain boundaries [55, 58-59]. At $m$ = 1/3, the SPS activation energy of $LnVO_4$ vanadates is lower than the typical activation energy of grain boundary diffusion in oxides and other compounds. As mentioned above, $LnVO_4$ vanadates are characterized by large unit cell volumes of the crystal lattice.

Let us discuss the reasons for the decrease in the SPS activation energy with an increase in the unit cell volume of the crystal lattice of vanadates (Fig. 21). The unit cell parameters increase monotonically with increasing Ln cation radius (Ln = La, Nd, Sm, Eu, Gd, Dy, Yb), which is located in the dodecahedral site of the vanadate crystal lattice. An increase of the unit cell volume of the vanadate crystal lattice can facilitate the diffusion of oxygen and metal ions in it, in its turn, an increase in the diffusion coefficients of oxygen and metal ions in the vanadate crystal lattice leads to an increase in the powder shrinkage intensity and a decrease of the sintering activation energy. A similar assumption could be made for diffusion along the grain boundaries. We emphasize that, the compounds EuVO$_4$, SmVO$_4$, GdVO$_4$ and DyVO$_4$ are close to metals, which can also make an additional contribution to the acceleration of their sintering under conditions of application of mechanical pressure. Similar effect is observed in tungsten carbide, which has abnormally low sintering and creep values [58, 59, 65].

**Conclusions**

1. Vanadates LnVO$_4$ and CaLnZr(VO$_4$)$_3$, where (Ln = La, Nd, Sm, Eu, Gd, Dy, Yb) were synthesized using the urea melt precipitation method. According to X-ray phase analysis, LaVO$_4$ vanadate crystallizes in the monazite structural type (analog, space group P2$_1$/n), the remaining vanadates belong to the zircon structural type (analog, space group I4$_1$/amd). The unit cell parameters increase monotonically with increasing radius of the cation in the dodecahedral site.

2. High-temperature *in-situ* XRD studies showed the maintenance of the phase composition up to 900°C. Based on the temperature dependences of the crystallographic parameters, the axial and average coefficients of thermal expansion (CTE) were calculated. The minimum CTE value was obtained for SmVO$_4$ ($\alpha_{av}$ = 5.49·10$^{-6}$ K$^{-1}$); the maximum is for LaVO$_4$ ($\alpha_{av}$ = 10.5·10$^{-6}$ K$^{-1}$).

3. Ceramic samples were prepared using the method of SPS from synthesized powders. Sintering was carried out by heating to a temperature of 600 °C at a rate of 100 °C/min, and then by heating to a temperature of 800 °C at a rate of 50 °C/min with no holding at the sintering temperature. The average duration of the sintering process was 8.1-8.3 min, excluding the duration of the cooling. The relative density of LnVO$_4$ ceramic samples did not exceed 78%, and CaLnZr(VO$_4$)$_3$ ceramic samples did not exceed 65%. The relative density of La$_{0.3}$Nd$_{0.5}$Sm$_{0.1}$Eu$_{0.1}$VO$_4$ ceramics was 40%. The phase composition and structural parameters of the ceramics correspond to the parameters of the original powders. It has been shown that the value of the effective sintering energy of vanadates correlates with the volume of the unit cell of the vanadate crystal lattice—compounds with a large unit cell volume V have a low SPS activation energy.

4. The hydrolytic stability of ceramic vanadates at room temperature was studied. In the static test mode, after 28 days, the minimum leaching rate for lanthanides reaches $0.04 \cdot 10^{-8}$ g·cm$^{-2}$·d$^{-1}$; for vanadium – $0.02 \cdot 10^{-5}$ g·cm$^{-2}$·d$^{-1}$. The maximum value of the leaching rate after 28 days of testing for lanthanide is $0.31 \cdot 10^{-8}$ g·cm$^{-2}$·d$^{-1}$, for vanadium – $1.56 \cdot 10^{-5}$ g·cm$^{-2}$·d$^{-1}$. This meets the requirements for materials for radioactive waste disposal. The resulting ceramic materials can be used to solve problems of HLW immobilization.

**Conflict of interest**. The authors declare that they have no known competing financial interests or personal relationships that could have appeared to influence the work reported in this paper.

**Author contribution statement**

**A.K. Koryttseva** – Conceptualization, Methodology, Analysis of experimental results, Writing of manuscript; **A.I. Orlova** – Project administration, Supervision, Funding acquisition, Writing of manuscript; **N.S. Litonova** – Investigation (Synthesis, XRD), Data curation, **A.V. Nokhrin** – Formal analysis, Analysis of experimental results, Writing - review & editing, Writing of manuscript, Data curation, **M.S. Boldin** – Investigation (SPS); **A.A. Murashov** & **D.G. Fukina** – Investigation (SEM, EDS); **A.A. Atopshev** – Investigation (Hydrolytic test); **K.E. Smetanina** – Investigation, Data curation (XRD); **A.I. Beskrovnyi** & **V.A. Turchenko** – Investigation (in-situ XRD); **N.Yu. Tabachkova** – Investigation (TEM).

**Acknowledgements**

The study was supported by the Russian Science Foundation (Grant #21-13-00308).

TEM study of the powders was carried out using the equipment of the Center for Collective Use "Materials Science and Metallurgy" (National University of Science and Technology "MISIS",

Moscow, Russia) with the financial support of the Ministry of Science and Higher Education of the Russian Federation (Grant #075-15-2021-696).

**Appendix A**

Ceramic samples ⌀ 10.8 mm were reheated to increase the density of ceramic vanadates, which characteristics are presented in Table. 3. Heating was carried out to a temperature of 1000 °C, the remaining heating modes (heating rate, applied pressure, holding time at sintering temperature) corresponded to the modes reported in Table 3.

Studies have shown that repeated heating up to 1000 °C leads to an increase in the relative density of ceramic vanadates by ~15-20%. The density of ceramic vanadates $DyVO_4$ and $NdVO_4$ increased to 92.53% (5.308 g/cm$^3$) and 93.05% (4.646 g/cm$^3$), respectively. The density of vanadates $CaDyZr(VO_4)_3$ and $CaEuZr(VO_4)_3$ increased to 75.63% (4.660 g/cm$^3$) and 82.22% (4.820 g/cm$^3$), respectively. Most of the samples were destroyed during reheating and their density cannot be reliably measured.

After heating up to 1000 °C, ceramic samples of vanadates have a fairly uniform microstructure (Fig. A1). Almost all vanadates exhibit grain growth. Anomalous grain growth is observed (Fig. A1c) in the microstructure of $CaEuZr(VO_4)_3$ ceramics. Microstructure of $NdVO_4$ ceramics indicates large dark particles, apparently formed as a result of the decomposition of the main vanadate phase (Fig. A1a).

The XRD results show that an insignificant content of impurity phases appears in ordinary vanadates $LnVO_4$ upon reheating. For example, when $DyVO_4$ ceramics were reheated, the appearance of the $DyVO_3$ phase was detected by XRD, but the intensity of its peaks only slightly exceeded the intensity of the diffraction pattern background. No increase of impurity phases amount was detected for some $LnVO_4$ vanadates ($NdVO_4$, $GdVO_4$, $EuVO_4$), within the accuracy of the X-ray diffraction method.

The appearance of a large number of extrinsic phases is detected for all CaLnZr(VO$_4$)$_3$ vanadates, after reheating. For example, GdVO$_4$ and mon-ZrO$_2$ phases was discovered in CaGdZr(VO$_4$)$_3$ ceramics, DyVO$_4$ and the zirconium oxide phase in monoclinic and tetragonal modifications were discovered in CaDyZr(VO$_4$)$_3$ ceramics; EuVO$_4$ phases, monoclinic and tetragonal ZrO$_2$ phases were detected in CaEuZr(VO$_4$)$_3$ ceramics.

It is interesting to note that after reheating, the XRD maxima of the LnVO$_4$ phase at angles 2θ ~ 24° (200), 33° (112) and 49° (400) bifurcate (see Fig. A2). We suppose this phenomenon to be explained by two factors:

- the presence of two phases with the same structure, but different unit cell volumes due to local differences in composition;

- the appearance of another structural type, which is very close to zircon, but has a distortion leading to a decrease in the system to rhombic, during heating. A similar decrease in symmetry is known for terbium vanadate TbVO$_4$ (No. 01-089-8055) with a low-temperature orthorhombic phase (space group Fddd).

# List of Figures

**Figure 1.** Powder XRD patterns of vanadates LnVO$_4$, where Ln – La, La$_{0.3}$Nd$_{0.5}$Sm$_{0.1}$Eu$_{0.1}$, Nd, Sm, Eu, Gd, Dy, Yb.

**Figure 2.** Powder XRD patterns of ternary vanadates CaLnZr(VO$_4$)$_3$, where Ln – Nd, Sm, Eu, Gd, Dy, Yb.

**Figure 3.** Dependence of unit cell parameters from Ln$^{3+}$ ionic radii for CaLnZr(VO$_4$)$_3$, where Ln – Nd, Sm, Eu, Gd, Dy, Yb.

**Figure 4.** Dependence of unit cell parameters from Ln$^{3+}$ ionic radii for LnVO$_4$, where Ln – La, La$_{0.3}$Nd$_{0.5}$Sm$_{0.1}$Eu$_{0.1}$, Nd, Sm, Eu, Gd, Dy, Yb.

**Figure 5.** Electron microscopic images of the vanadates powders: CaYbZr(VO$_4$)$_3$ (a, b), CaNdZr(VO$_4$)$_3$ (c, d).

**Figure 6.** Electron microscopic images of the vanadates powders SmVO$_4$ (a, б) и GdVO$_4$ (в, г).

**Figure 7.** SEM-images (a, b) and TEM-images (c, d, e, f) of the surfaces of the vanadate powders CaYbZr(VO$_4$)$_3$, prepared from solution with the citric acid (a, c, d) and from molten urea (b, e, f).

**Figure 8.** TEM-images of powders YbVO$_4$ at different magnifications: (a) individual nanoparticles, (b) several particles, (c) particles agglomeration.

**Figure 9.** XRD patterns, obtained during direct heating of powders SmVO$_4$ from 30 to 900°C

**Figure 10.** XRD patterns, obtained during direct heating of the powder CaYbZr(VO$_4$)$_3$, prepared from solution with citric acid.

**Figure 11.** XRD patterns, obtained during direct heating of the powder CaNdZr(VO$_4$)$_3$: (a) overview diffraction patterns; (b) selected pattern at 900°C.

**Figure 12.** Temperature dependence of unit cell parameters for LnVO$_4$, where Ln – Nd, Sm, Eu, Gd, Dy, Yb.

**Figure 13.** Temperature dependence of unit cell parameters for CaLnZr(VO$_4$)$_3$, where Ln – Nd, Sm, Eu, Gd, Dy, Yb.

**Figure 14.** Typical diagram of ceramic sintering "Temperature (T) – Applied Pressure (P) – Vacuum Pressure (V$_{ac}$) – Time (t)" (for EuVO$_4$ ceramic)

**Figure 15.** SEM-images of ceramic surfaces and elements distribution maps (Ca, Zr, V, Yb, Sm, Nd) throughout the surface: SmVO$_4$ (a), NdVO$_4$ (b), CaNdZr(VO$_4$)$_3$ (c) and CaYbZr(VO$_4$)$_3$ (d)

**Figure 16.** Ununiformed microstructure in ceramics LaVO$_4$ (a), CaYbZr(VO$_4$)$_3$ (b) and CaSmZr(VO$_4$)$_3$ (c).

**Figure 17.** Time dependence of normalized mass loss (NL) and leaching rate (R) for different ceramic samples: *a* – La leaching from LaVO$_4$, *b* – Sm leaching from CaSmZr(VO$_4$)$_3$, *c* – V leaching from LaVO$_4$, *d* – V leaching from CaSmZr(VO$_4$)$_3$.

**Figure 18**. XRD patterns of selected samples before and after leaching test.

**Figure 19**. Temperature curves of effective shrinkage (a, b) and true shrinkage (c, d) of ceramic vanadates: (a, c) compounds LnVO$_4$; (b, d) compounds CaLnZr(VO$_4$)$_3$. L$_0$(T) – temperature curve of an empty graphite mold thermal expansion.

**Figure 20.** Analysis of thermal curves of shrinkage of vanadate powders during SPS. Dependence ln(T$\partial\varepsilon/\partial$T) – T$_m$/T for compound EuVO$_4$.

**Figure 21.** Dependence of effective SPS activation energy from vanadates unit cell volume. ○ – data for EuVO$_4$.

**Figure A1.** Microstructure of ceramic vanadate fractures NdVO$_4$ (*a*), DyVO$_4$ (*b*), CaEuZr(VO$_4$)$_3$ (*c*), CaDyZr(VO$_4$)$_3$ (*d*). Heating up to 1000 ºC.

**Figure A2.** Diffraction pattern of ceramics CaEuZr(VO$_4$)$_3$ (*a*), CaDyZr(VO$_4$)$_3$ (*b*) и CaGdZr(VO$_4$)$_3$ (*c*) after reheating

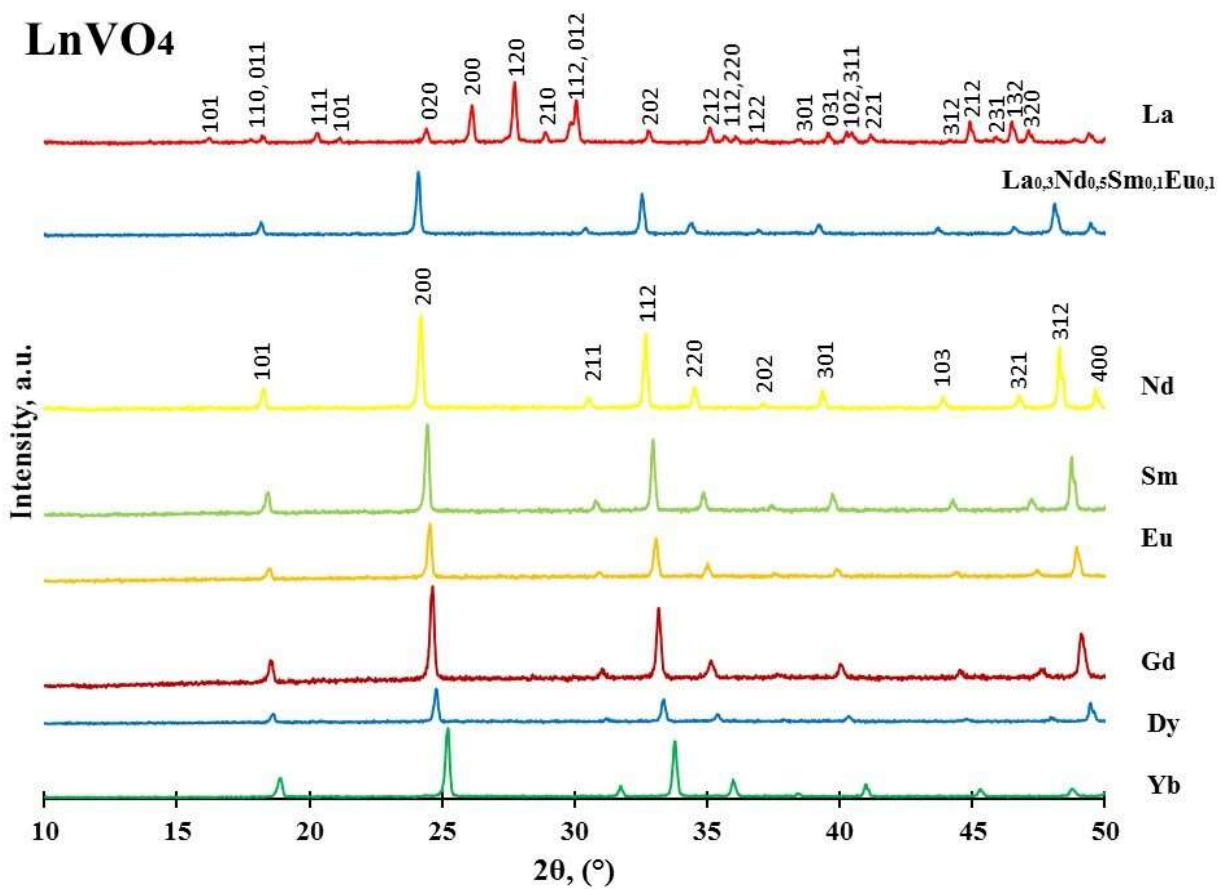

Figure 1

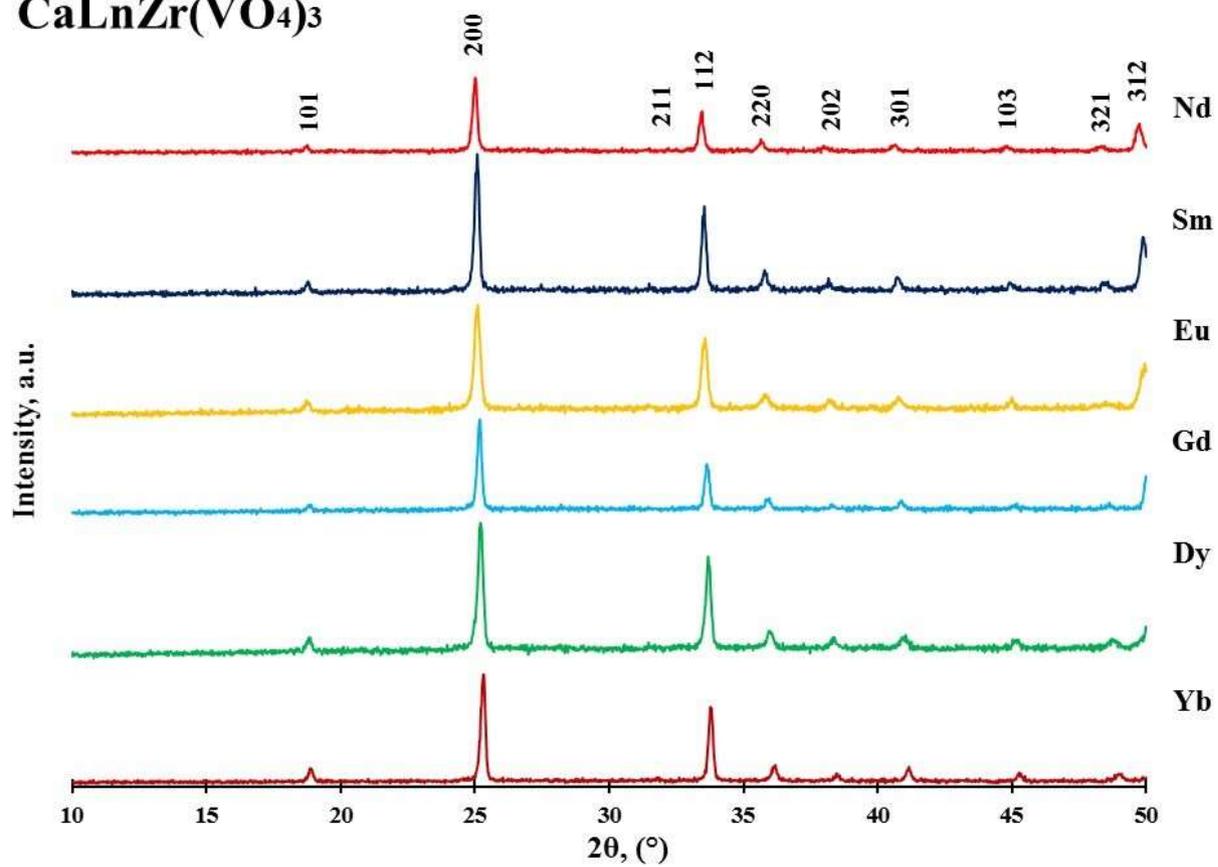

Figure 2

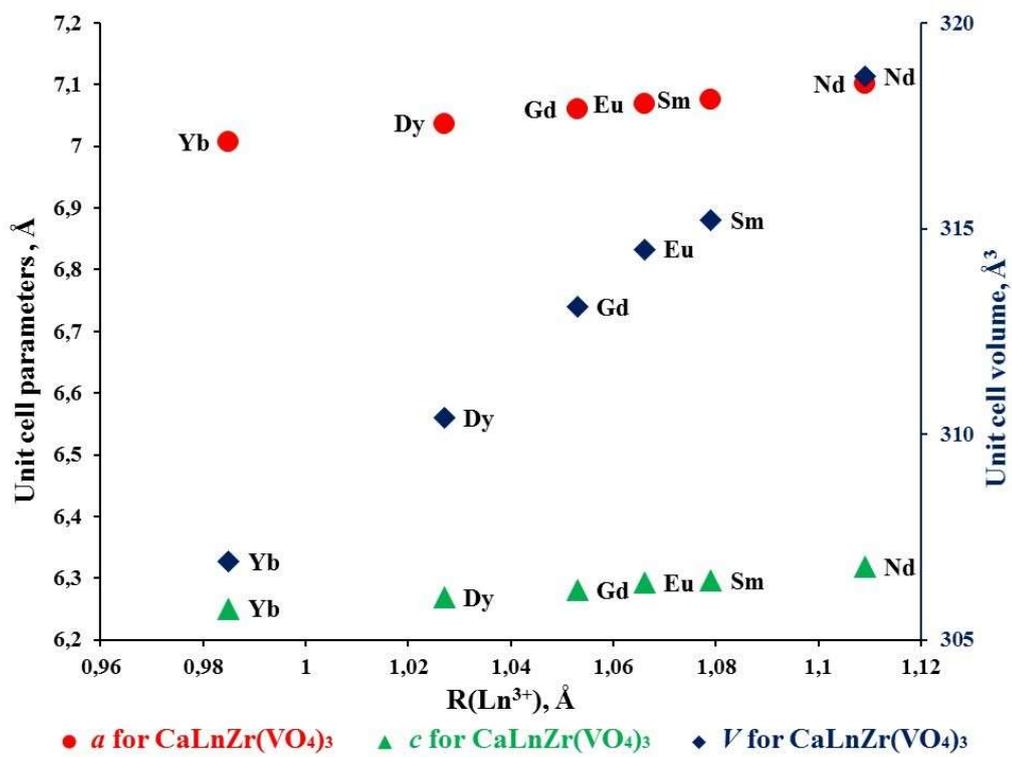

Figure 3

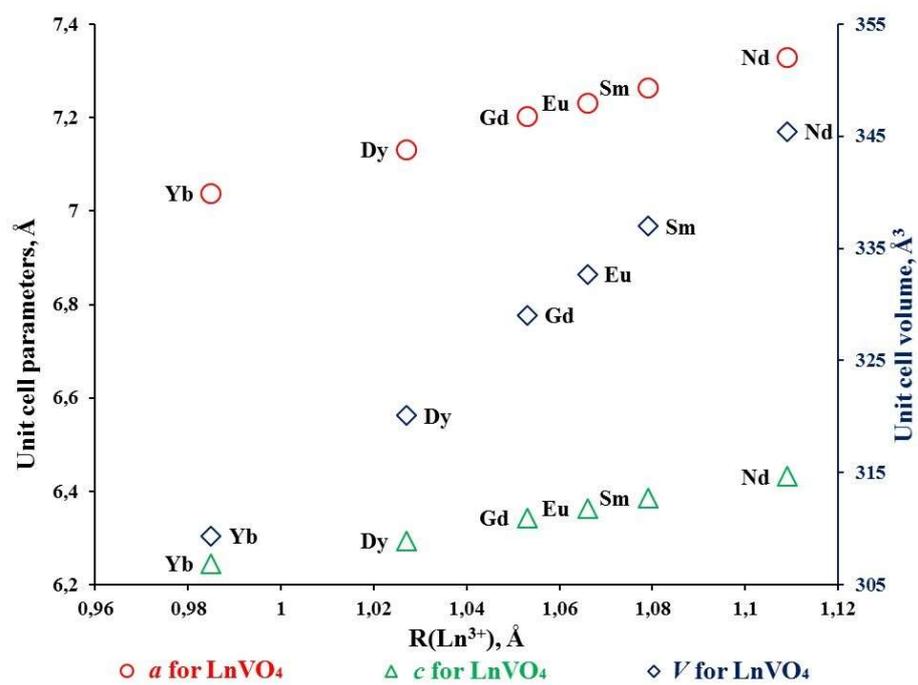

Figure 4

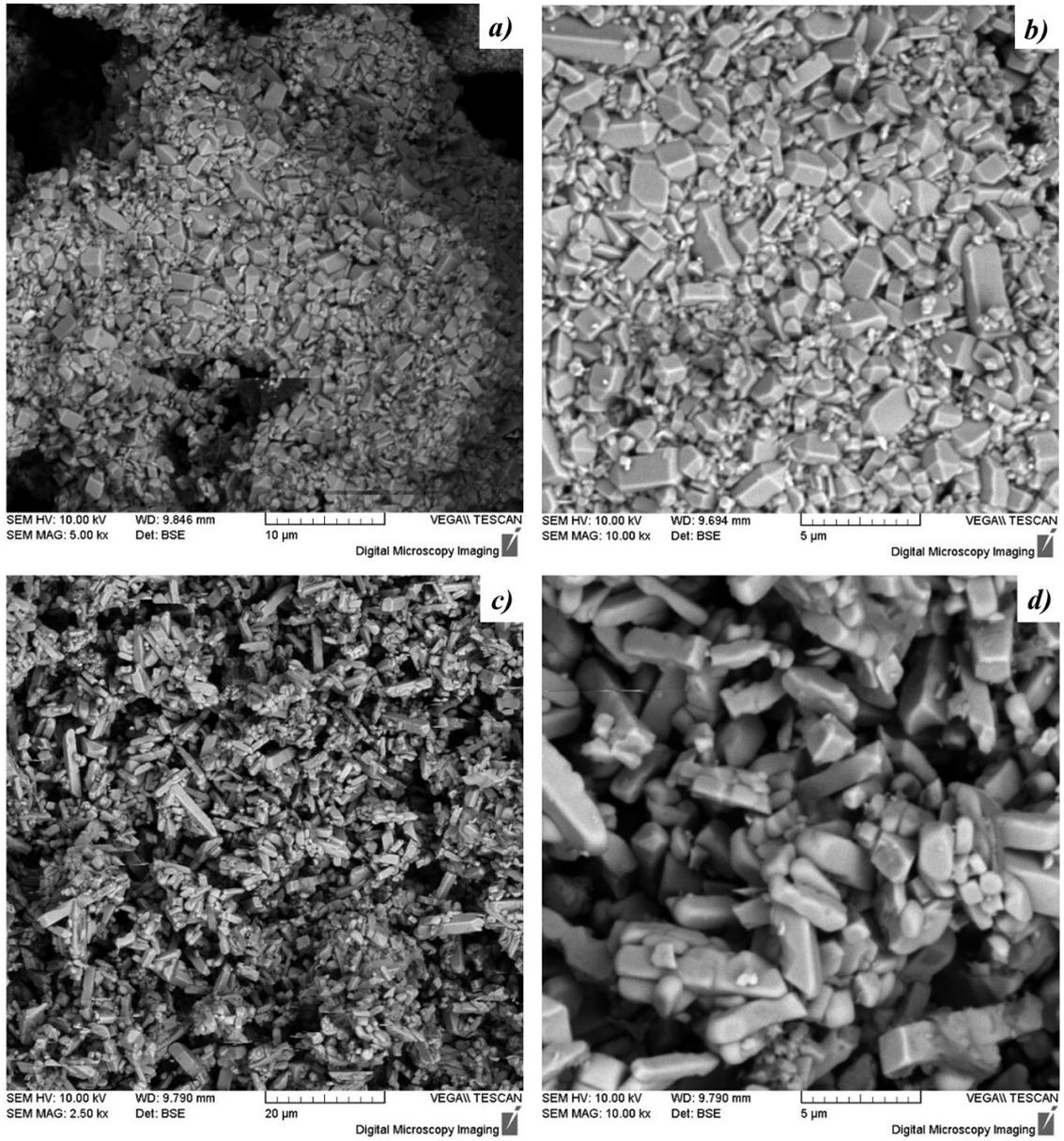

Figure 5

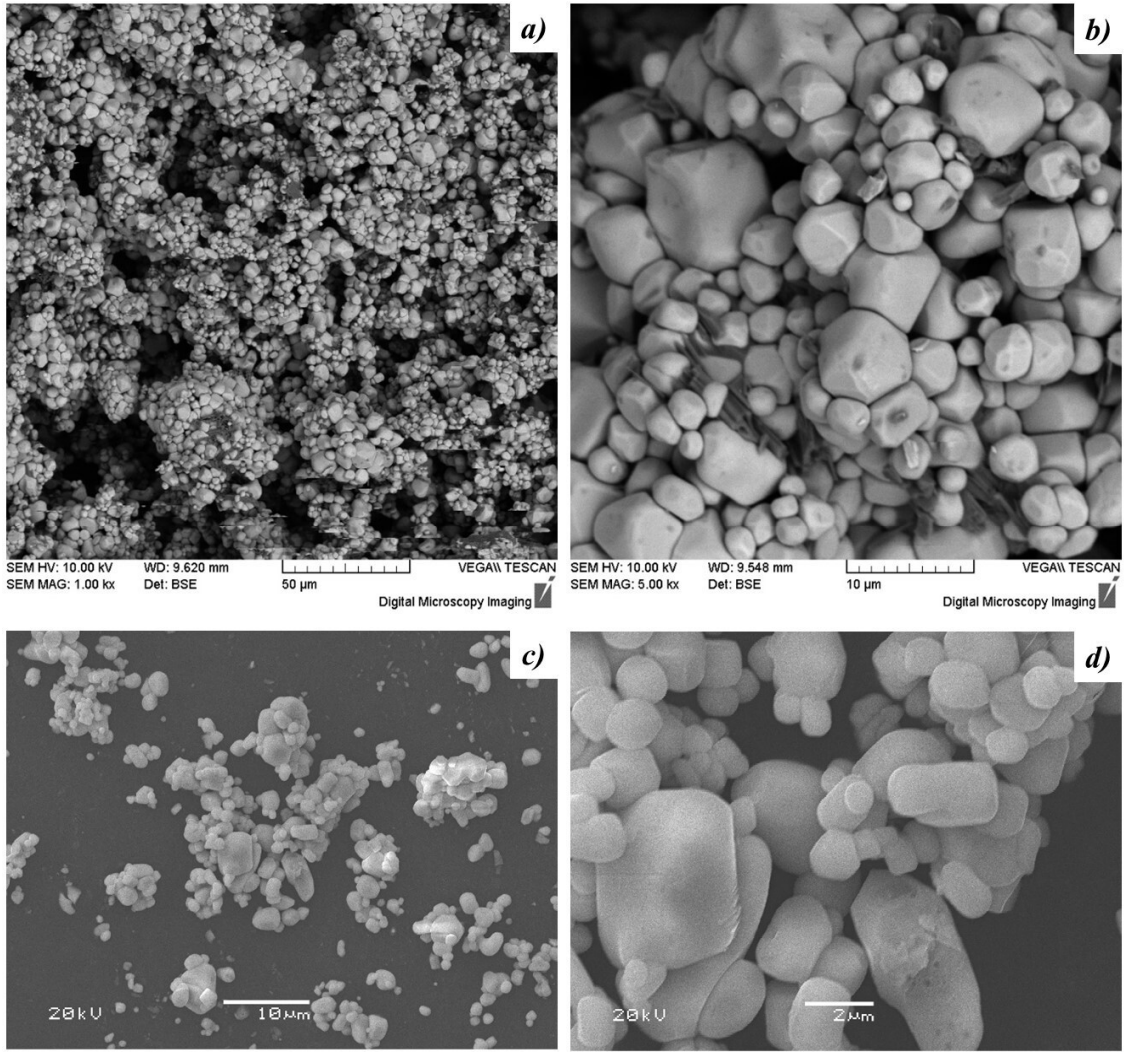

Figure 6

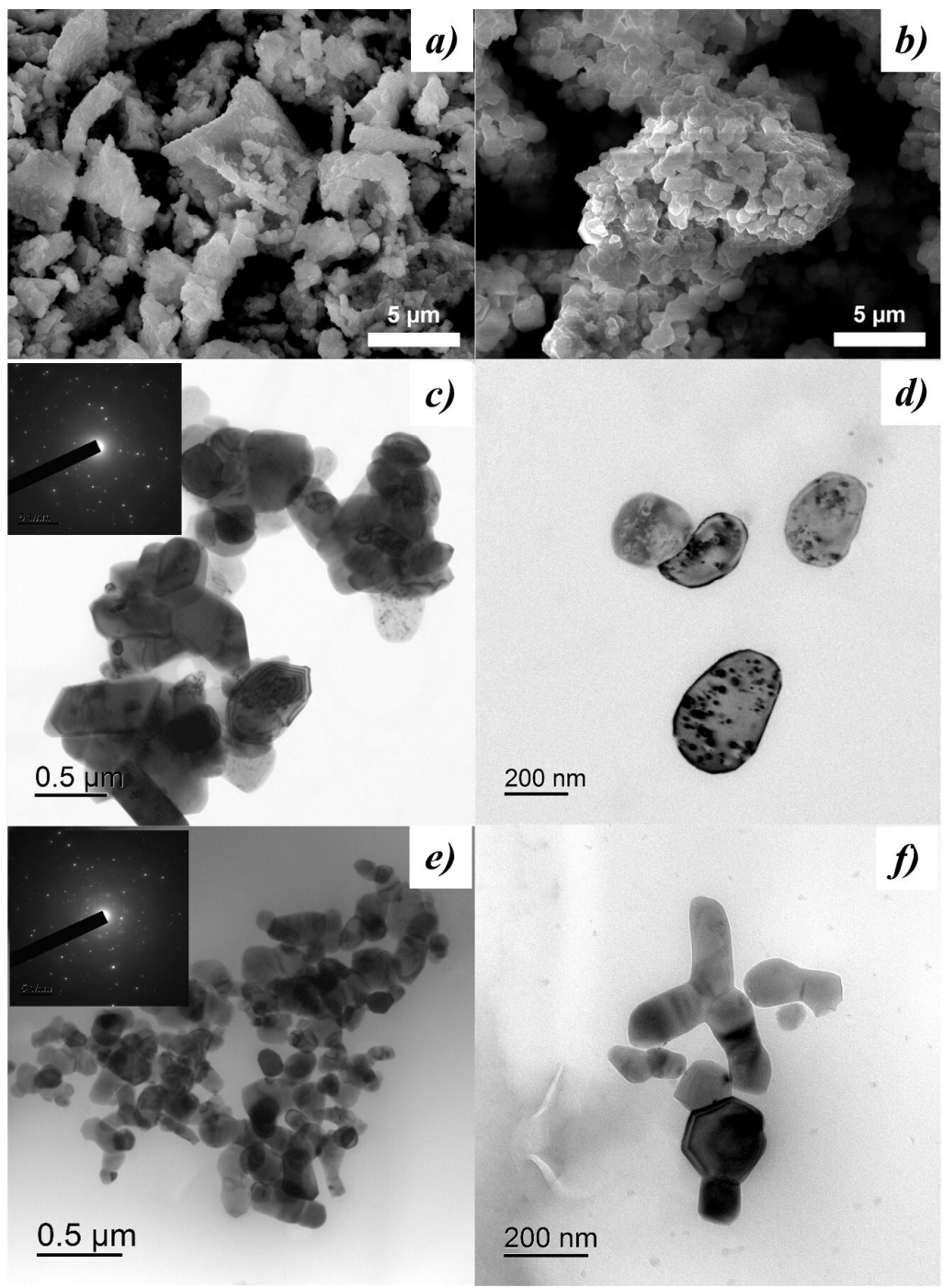

Figure 7

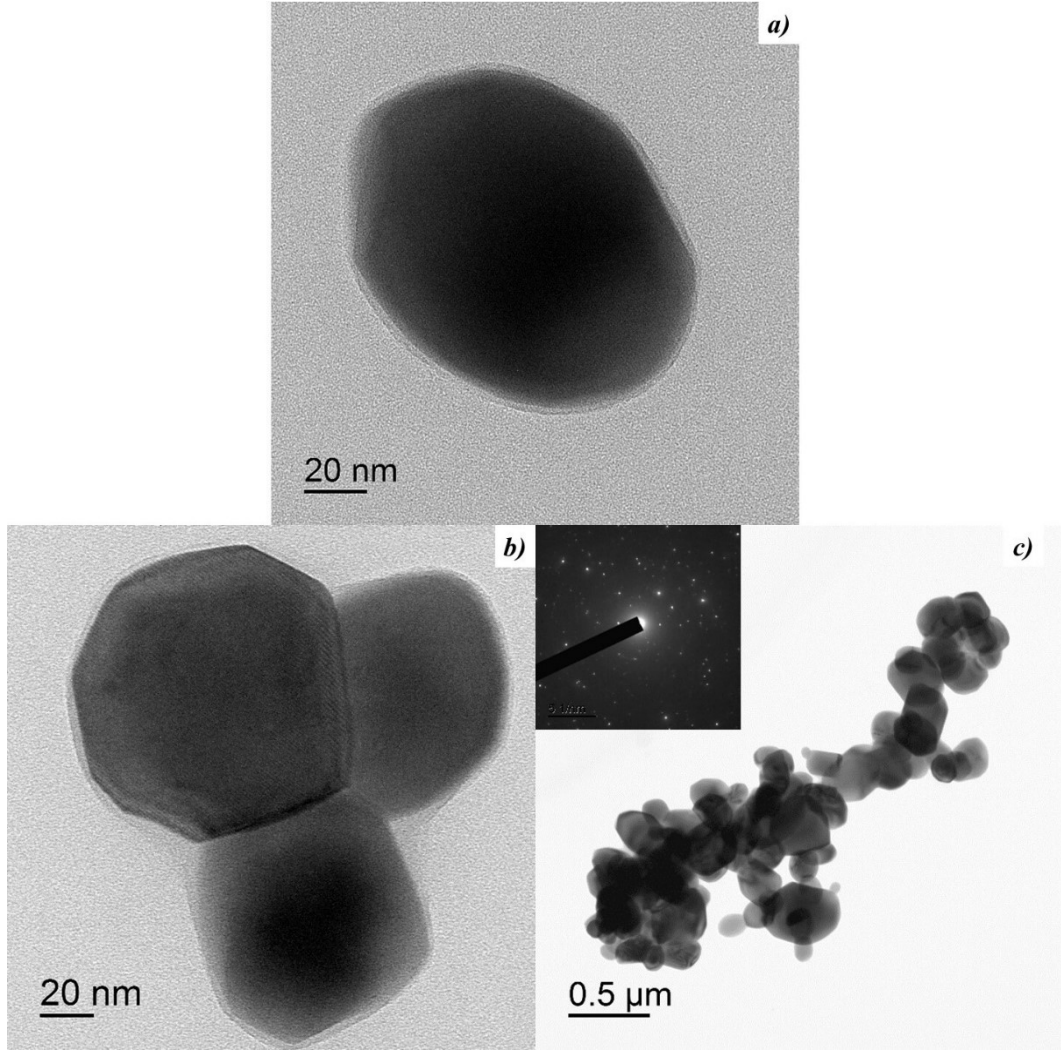

Figure 8

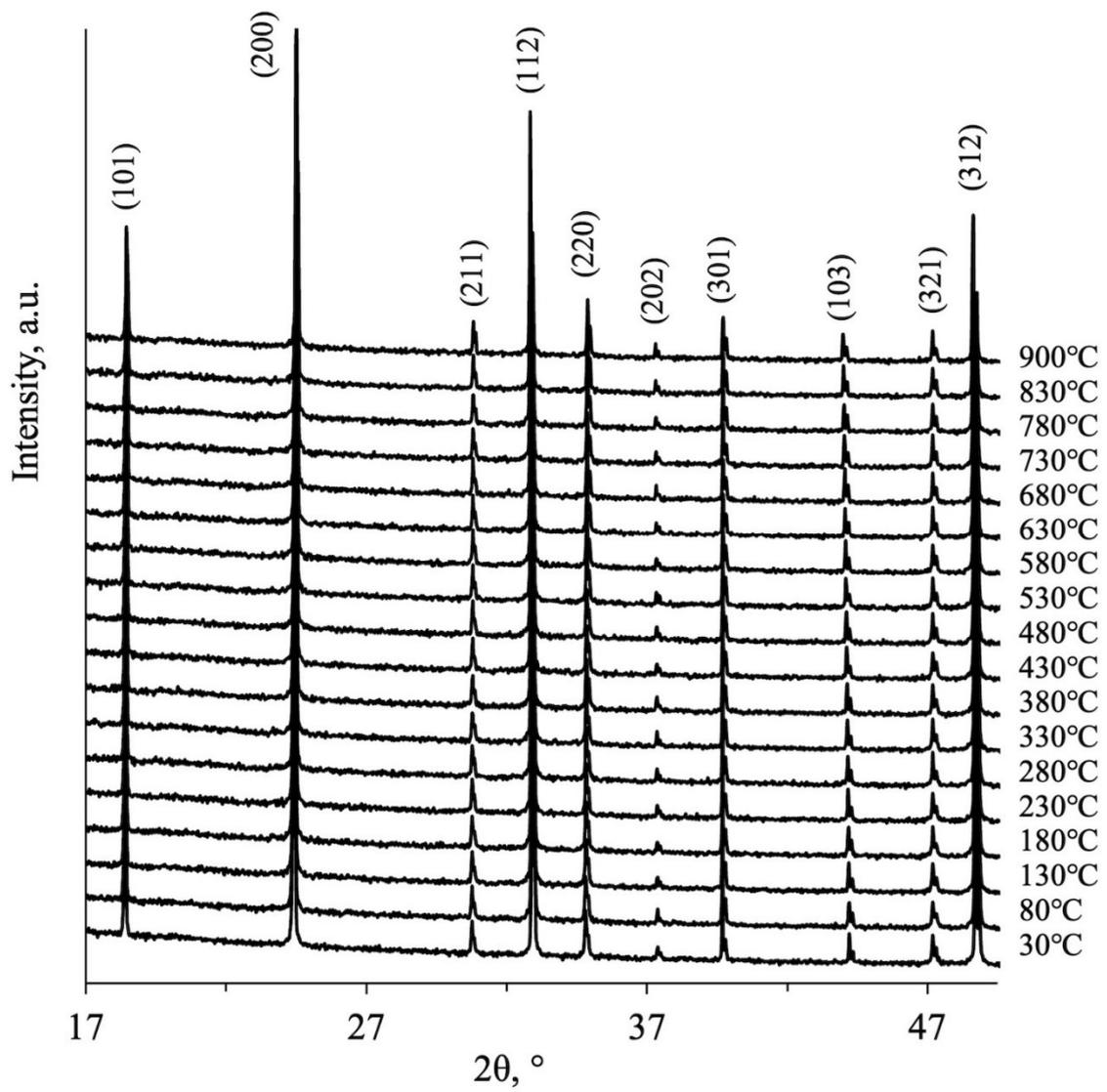

Figure 9

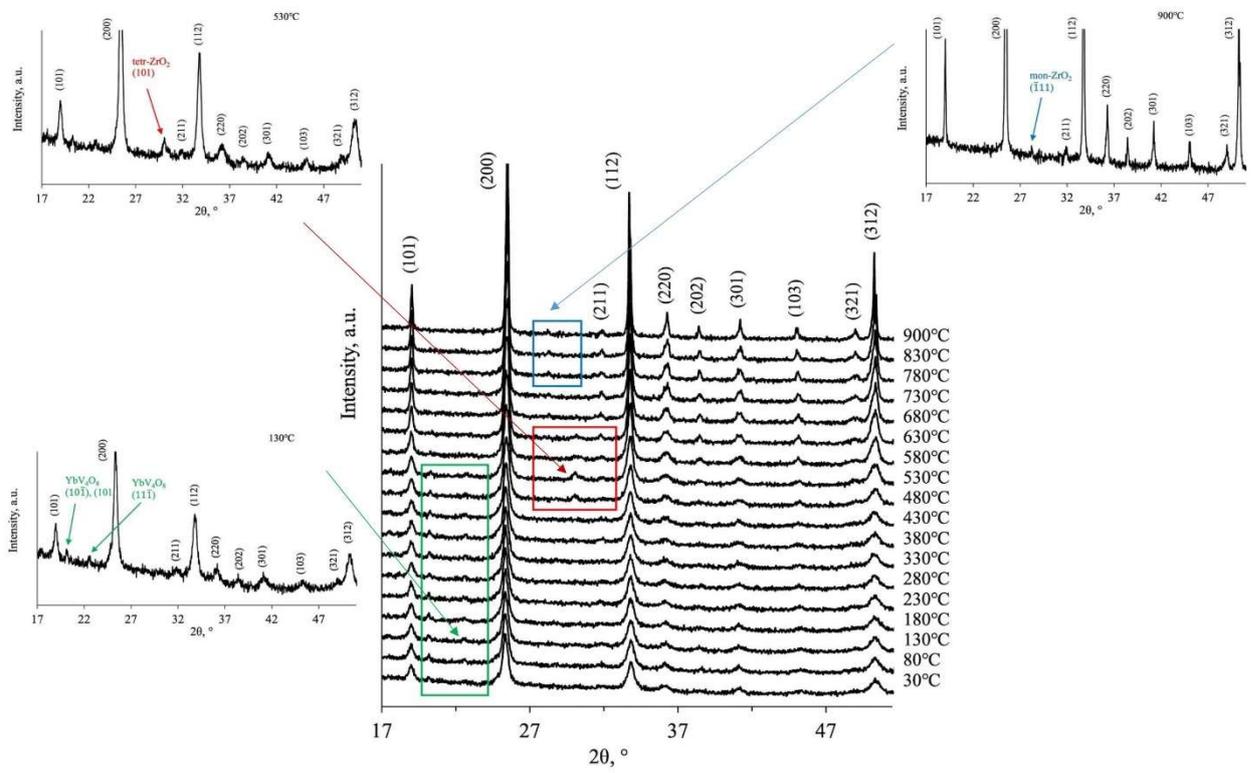

Figure 10

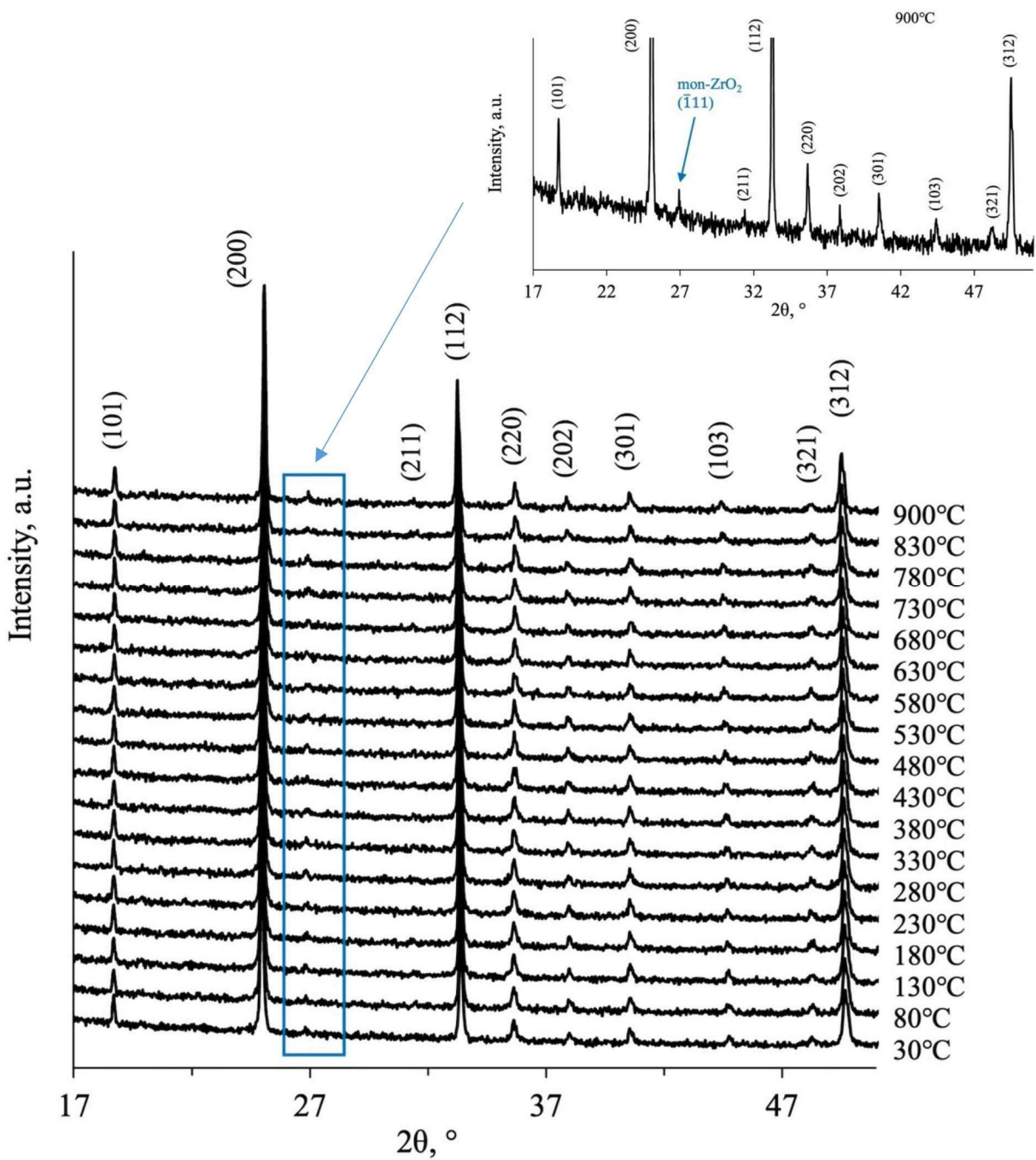

Figure 11

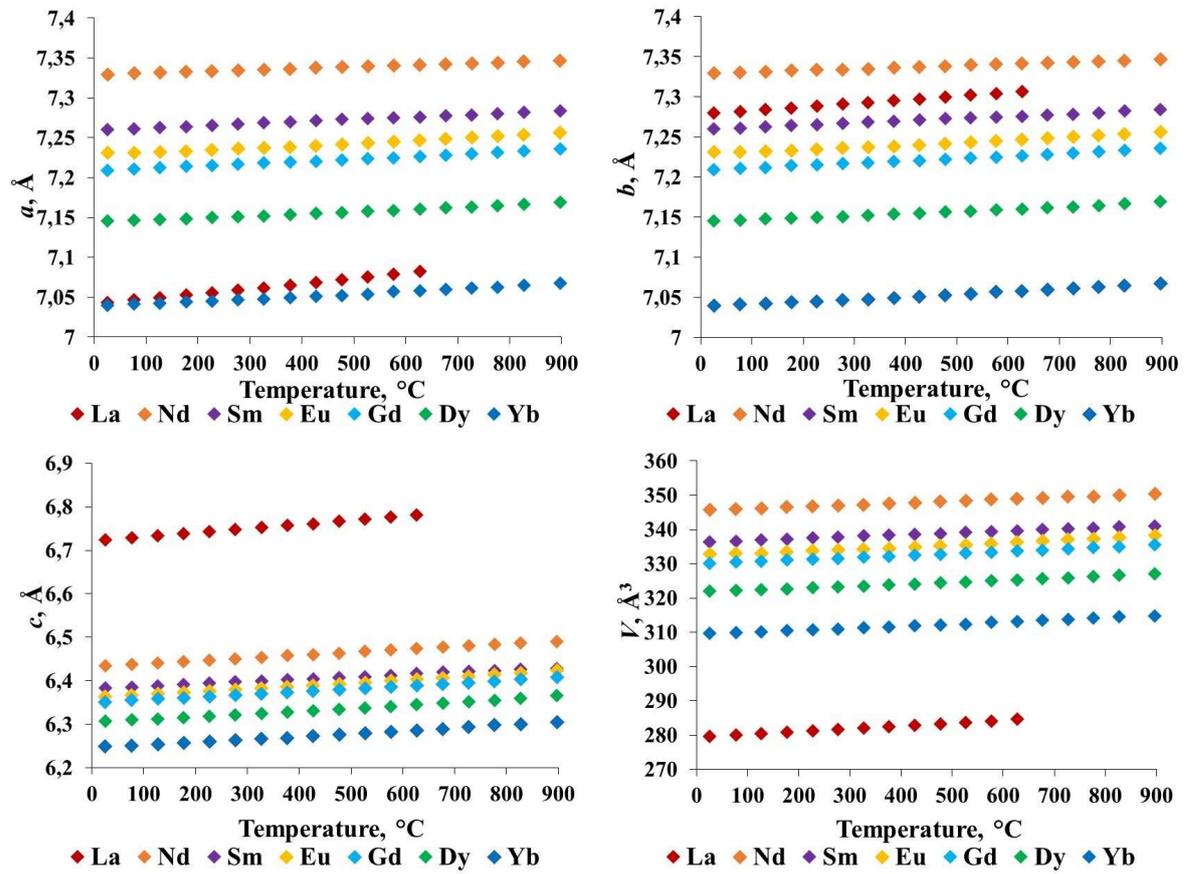

Figure 12

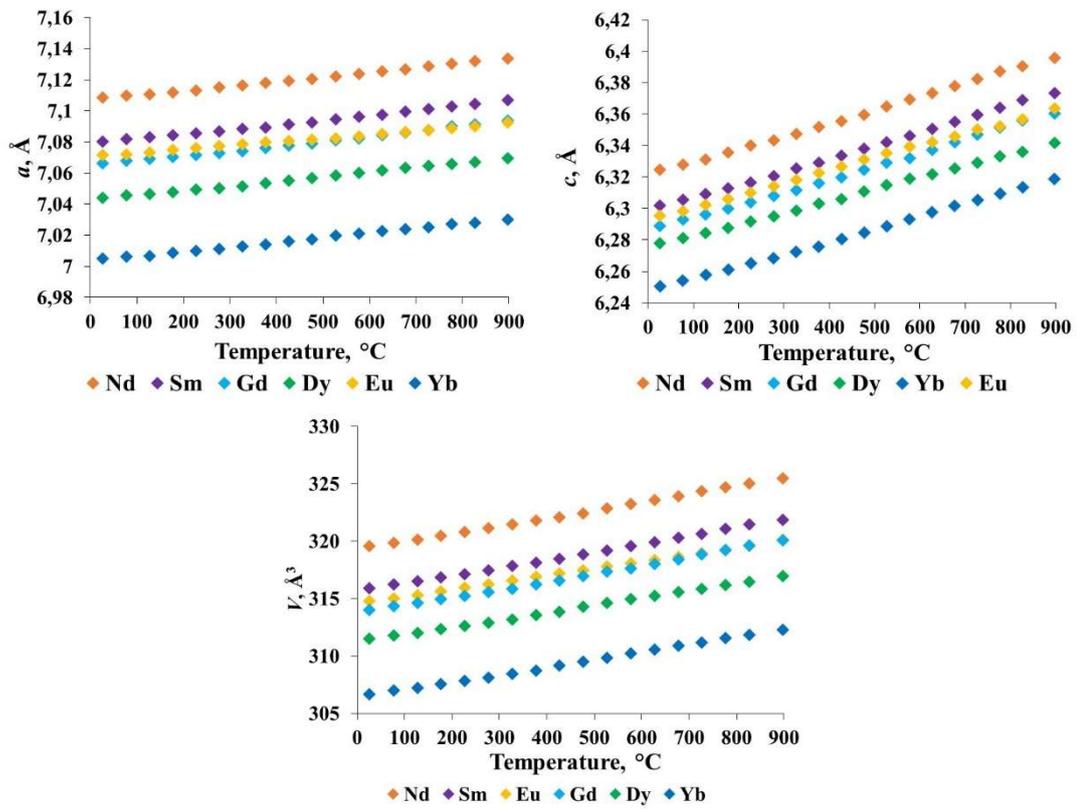

Figure 13

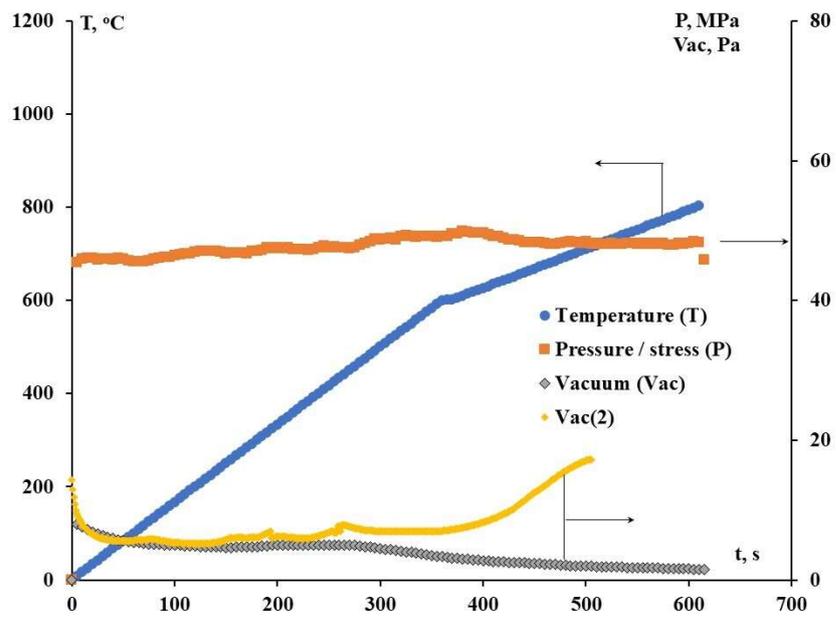

Figure 14

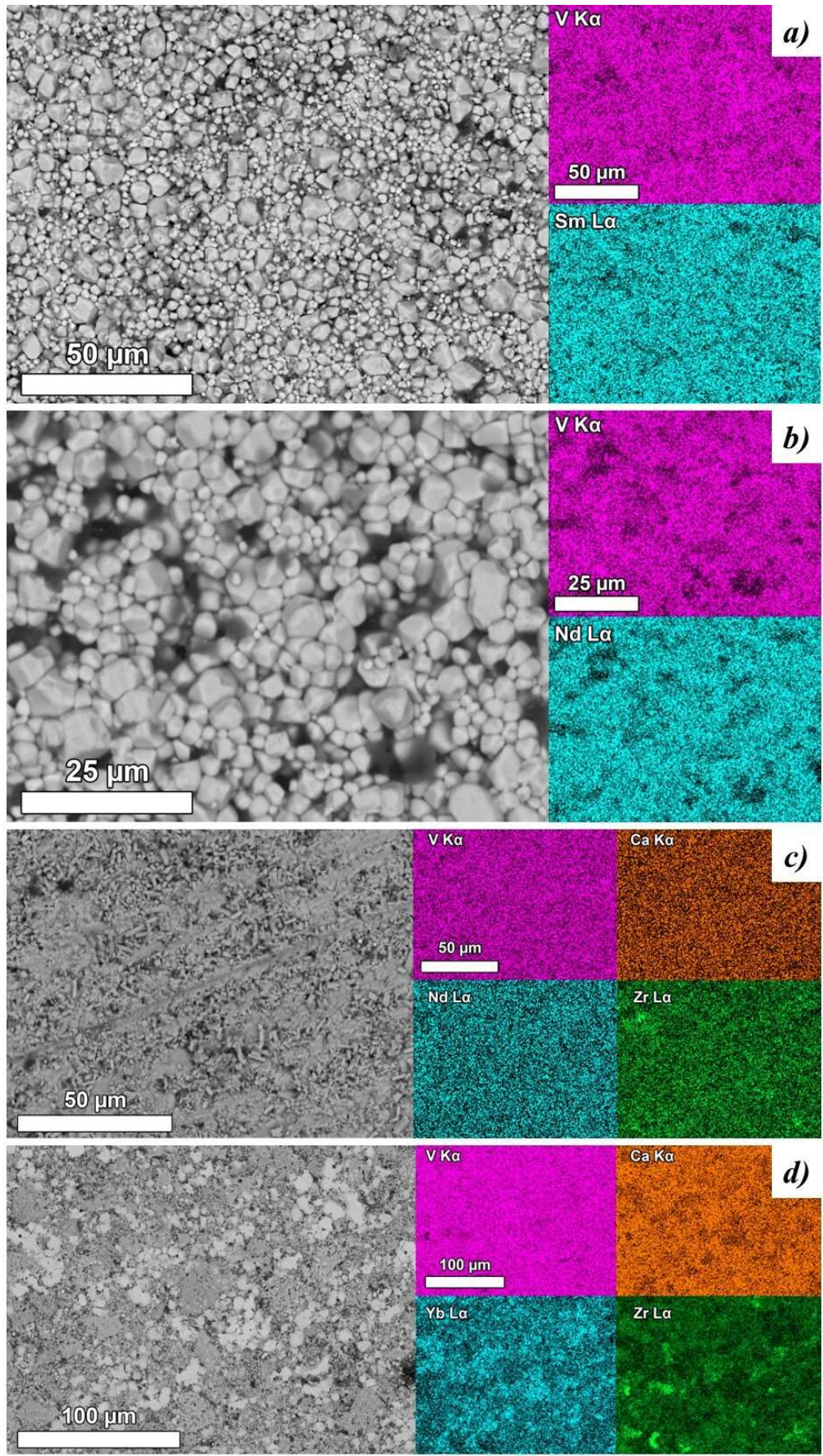

Figure 15

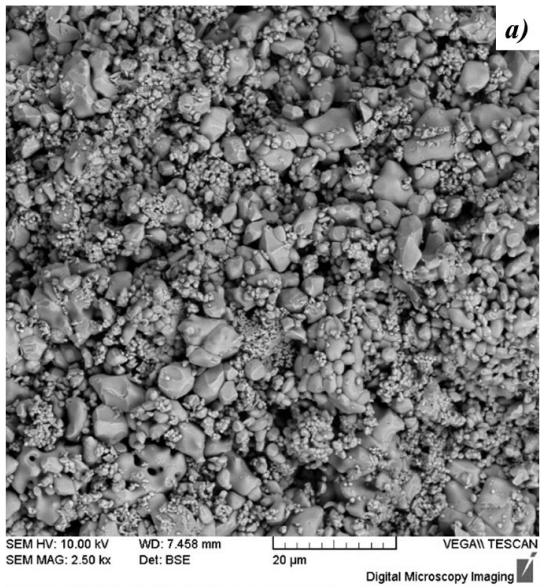
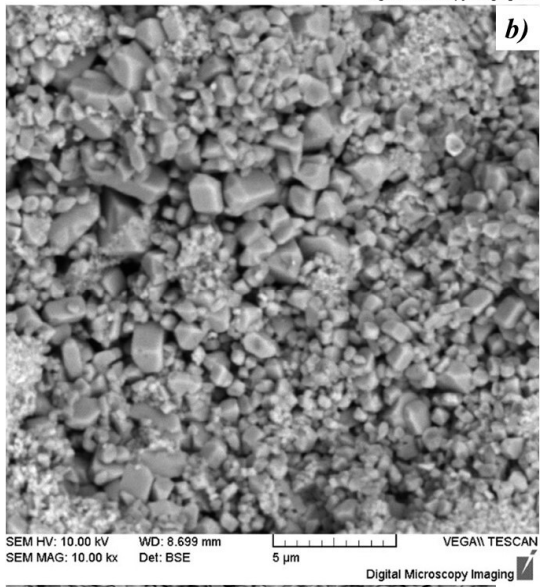
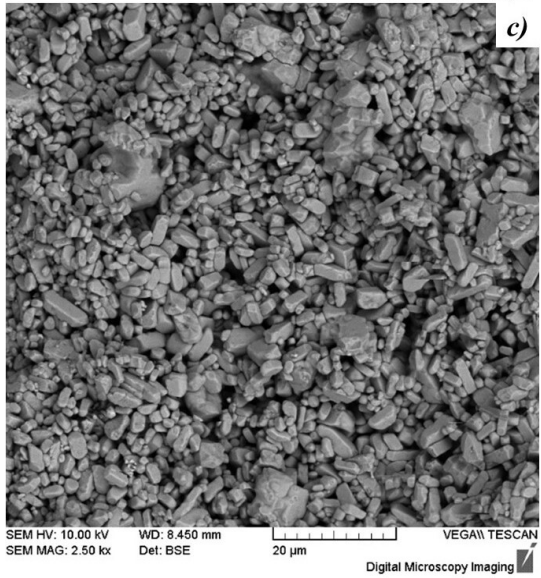

Figure 16

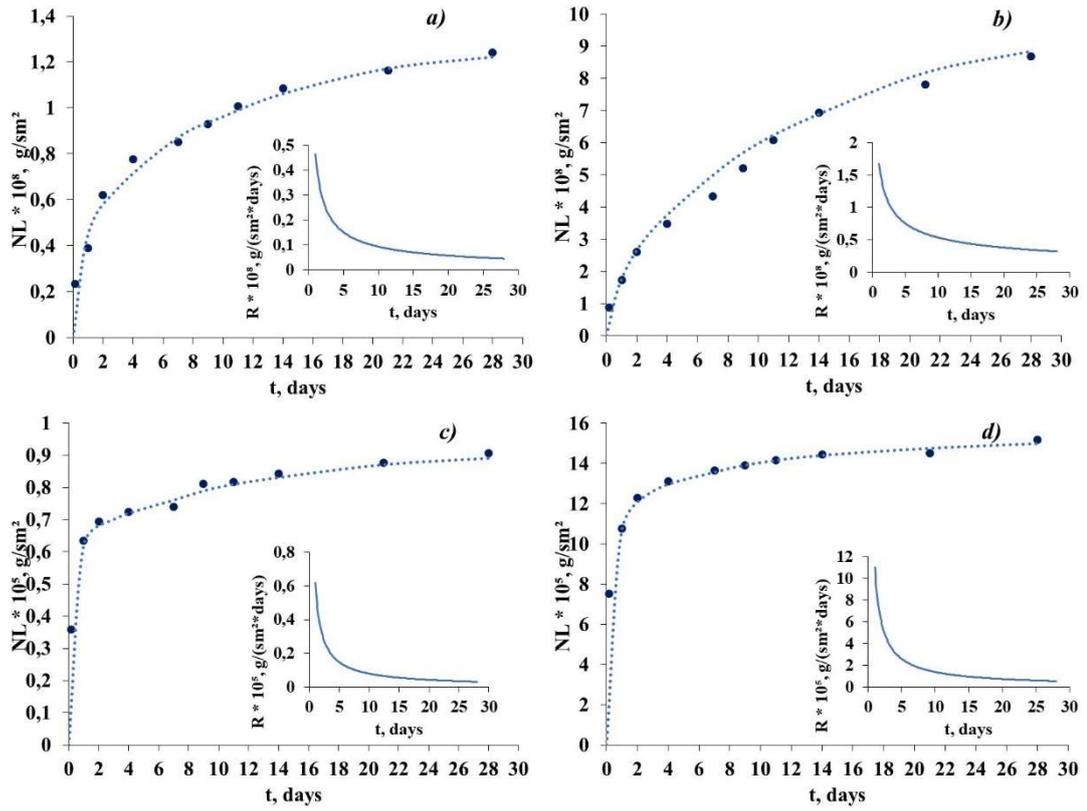

Figure 17

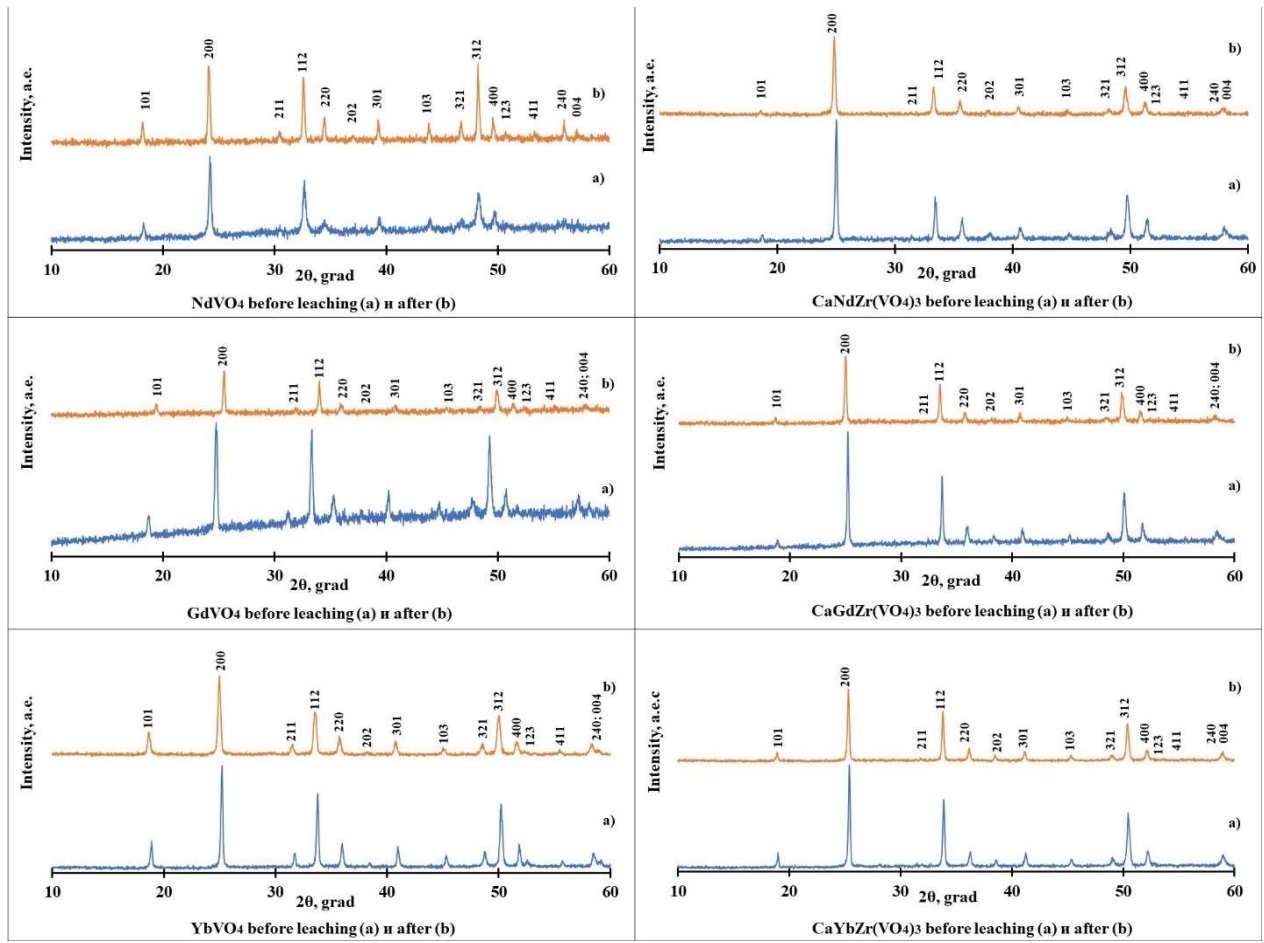

Figure 18

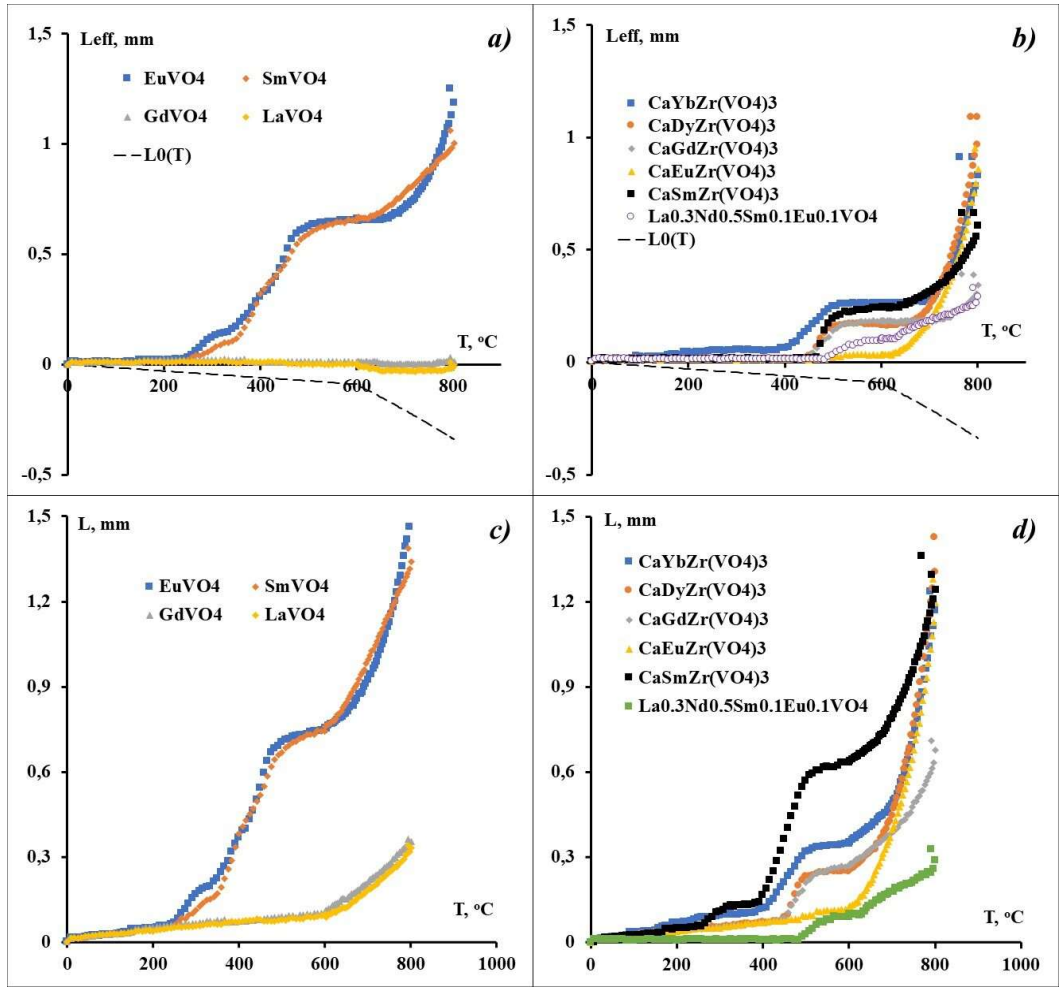

Figure 19

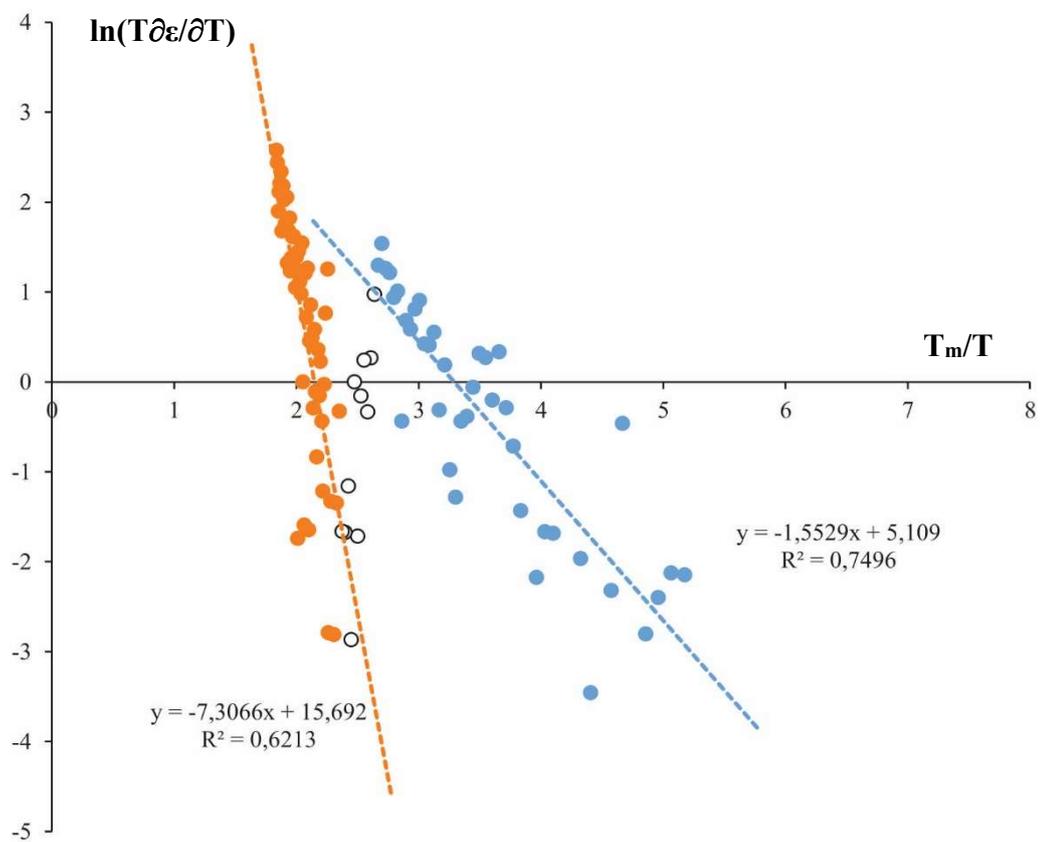

Figure 20

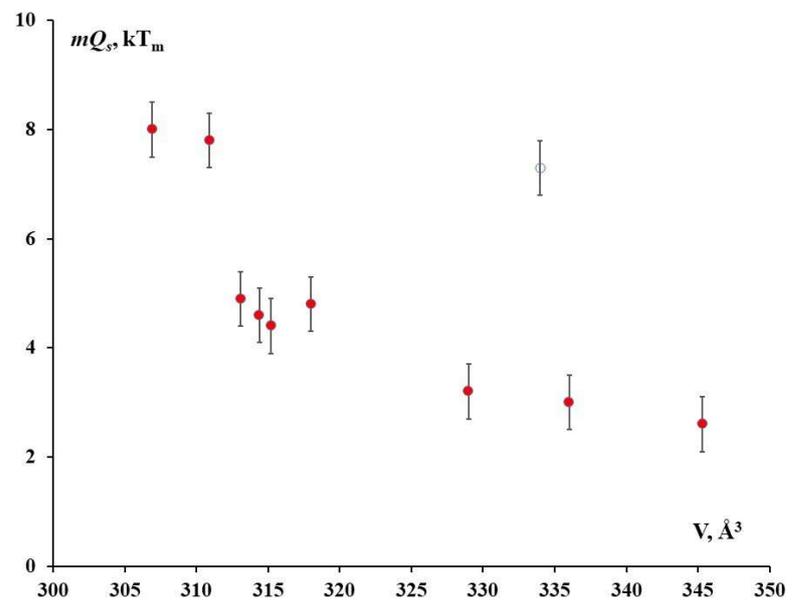

Figure 21

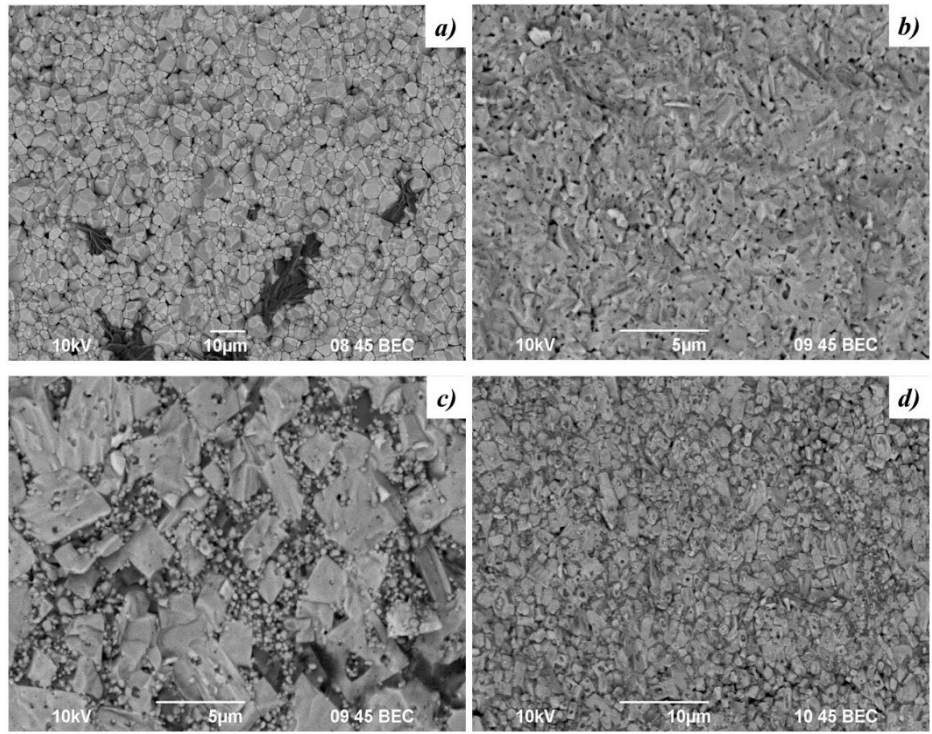

Figure A1

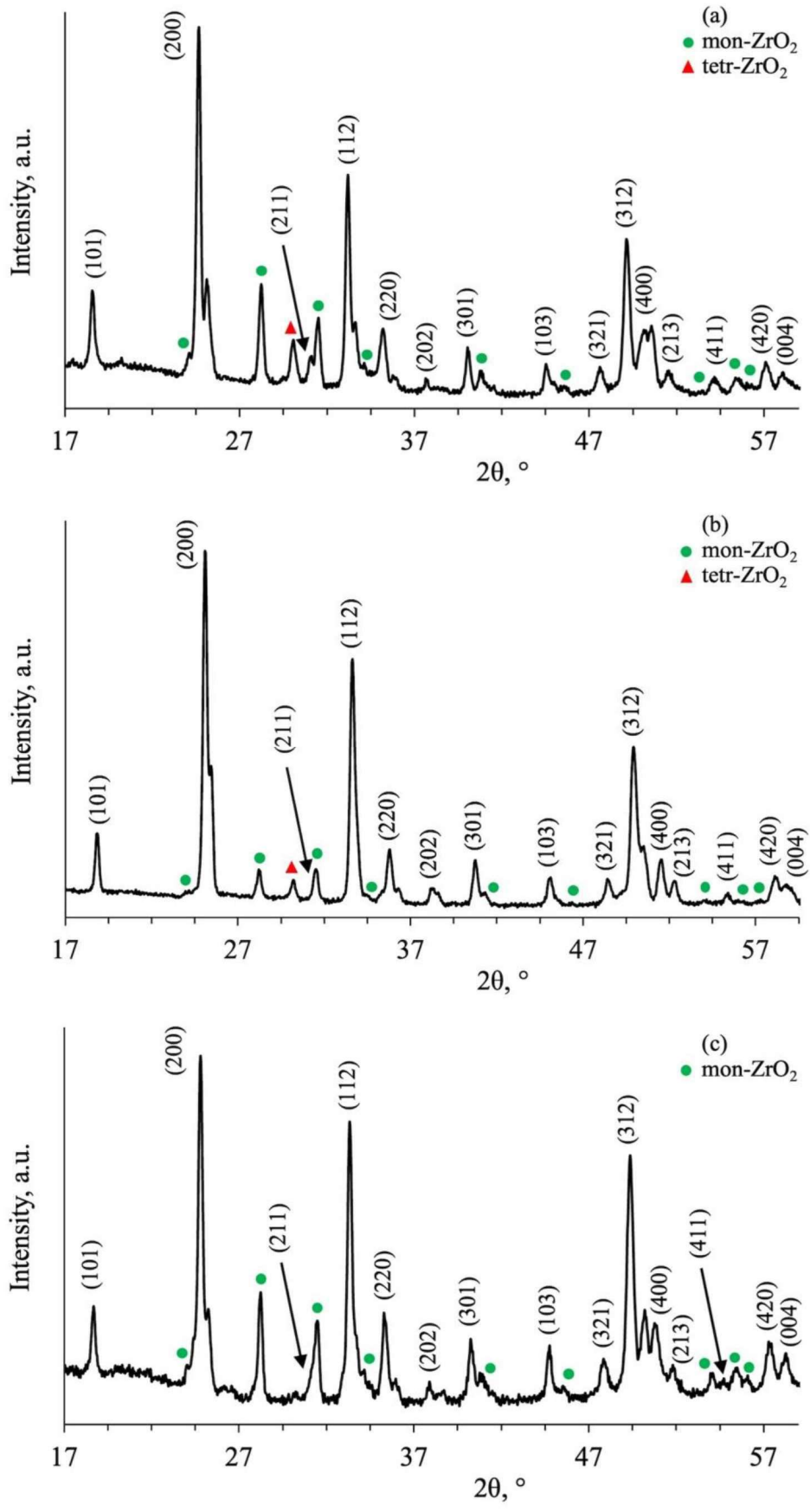

Figure A2